\documentclass[twocolumn,twocolappendix]{aastex631}
\pdfoutput=1
\usepackage{amsmath,amstext}
\let\tablenum\relax
\usepackage{siunitx}
\usepackage[T1]{fontenc}
\usepackage{times}
\usepackage{apjfonts} 
\usepackage[figure,figure*]{hypcap}
\usepackage{natbib}
\usepackage{enumerate}
\usepackage{booktabs}
\usepackage{graphicx}
\usepackage{multirow}
\usepackage[encapsulated]{CJK}
\usepackage[caption=false]{subfig}
\usepackage{hyperref}
\graphicspath{{./}{Figures/}}

\newcommand{\grizli}{\textsc{Gri$z$li}}
\newcommand{\eazy}{\textsc{EA$z$Y}}
\newcommand{\pros}{\textsc{Prospector}}
\newcommand{\prosa}{\pros{}-$\alpha$}

\newcommand{\galfit}{\textsc{GALFIT}}

\newcommand{\ninit}{9263}
\newcommand{\ninitgold}{2872}
\newcommand{\nphotrem}{963}
\newcommand{\nphotremfrac}{11.6}
\newcommand{\nspeconly}{7}
\newcommand{\ncmd}{1597}
\newcommand{\ncmdgold}{755}
\newcommand{\nuvj}{105}
\newcommand{\nuvjgold}{38}
\newcommand{\nmsrem}{167}
\newcommand{\nmsremfrac}{9.7}
\newcommand{\nnirspectot}{27}
\newcommand{\nnirspecadd}{14}
\newcommand{\ncluster}{1531}
\newcommand{\nclustergold}{761}

\newcommand{\ngalfail}{50}
\newcommand{\ngalfailfrac}{3.2}
\newcommand{\nflag}{313}
\newcommand{\nflagfrac}{20.2}
\newcommand{\nrff}{232}
\newcommand{\nrfffrac}{15}
\newcommand{\nlowsigma}{4}
\newcommand{\nlowsigmafrac}{0.3}
\newcommand{\nmasslim}{9}
\newcommand{\nmasslimfrac}{0.9}
\newcommand{\nhimass}{81}
\newcommand{\nfinal}{991}
\newcommand{\nfinalgold}{488}
\newcommand{\lmfrac}{6.5}
\newcommand{\hmfrac}{43.5}

\newcommand{\alphaval}{0.206}
\newcommand{\alphaerr}{0.016}
\newcommand{\logaval}{0.565}
\newcommand{\logaerr}{0.040}
\newcommand{\logsreval}{0.193}
\newcommand{\logsreerr}{0.005}
\newcommand{\logsrevalcirc}{0.181}
\newcommand{\logsreerrcirc}{0.005}

\received{2025 April 14}
\revised{2025 August 12}
\accepted{XXX}
\submitjournal{the Astrophysical Journal}
\shorttitle{Low-Mass Quiescent Galaxies in Abell 2744}
\shortauthors{Cutler et al.}

\begin{document}
\title{The Structure and Formation Histories of Low-Mass Quiescent Galaxies in the Abell 2744 Cluster Environment}

\correspondingauthor{Sam E. Cutler}
\email{secutler@umass.edu}

\author[0000-0002-7031-2865]{Sam E. Cutler}
\affiliation{Department of Astronomy, University of Massachusetts, Amherst, MA 01003, USA}

\author[0000-0003-1614-196X]{John R. Weaver}
\affiliation{Department of Astronomy, University of Massachusetts, Amherst, MA 01003, USA}

\author[0000-0001-7160-3632]{Katherine E. Whitaker}
\affiliation{Department of Astronomy, University of Massachusetts, Amherst, MA 01003, USA}
\affiliation{Cosmic Dawn Center (DAWN), Denmark}

\author[0000-0002-5612-3427]{Jenny E. Greene}
\affiliation{Department of Astrophysical Sciences, Princeton University, 4 Ivy Ln., Princeton, NJ 08544, USA}

\author[0000-0003-4075-7393]{David J. Setton}
\thanks{Brinson Prize Fellow}
\affiliation{Department of Astrophysical Sciences, Princeton University, 4 Ivy Ln., Princeton, NJ 08544, USA}

\author[0009-0002-6073-8026]{Zach J. Webb}
\affiliation{Department of Astronomy, University of Massachusetts, Amherst, MA 01003, USA}

\author[0009-0002-5707-2809]{Ayesha Abdullah}
\affiliation{Department of Astronomy, University of Massachusetts, Amherst, MA 01003, USA}

\author[0000-0002-8450-9992]{Aubrey Medrano}
\affiliation{Department of Astronomy, University of Massachusetts, Amherst, MA 01003, USA}

\author[0000-0001-5063-8254]{Rachel Bezanson}
\affiliation{Department of Physics and Astronomy and PITT PACC, University of Pittsburgh, Pittsburgh, PA 15260, USA}

\author[0000-0003-2680-005X]{Gabriel Brammer}
\affiliation{Cosmic Dawn Center (DAWN), Denmark}
\affiliation{Niels Bohr Institute, University of Copenhagen, Jagtvej 128, 2200 Copenhagen N, Denmark}

\author[0000-0002-1109-1919]{Robert Feldmann}
\affiliation{Department of Astrophysics, University of Zurich, Zurich CH-8057, Switzerland}

\author[0000-0001-6278-032X]{Lukas J. Furtak}
\affiliation{Physics Department, Ben-Gurion University of the Negev, P.O. Box 653, Be’er-Sheva 84105, Israel}

\author[0000-0002-3254-9044]{Karl Glazebrook}
\affiliation{Centre for Astrophysics and Supercomputing, Swinburne University of Technology, PO Box 218, Hawthorn, VIC 3122, Australia}

\author[0000-0002-2057-5376]{Ivo Labbe}
\affiliation{Centre for Astrophysics and Supercomputing, Swinburne University of Technology, PO Box 218, Hawthorn, VIC 3122, Australia}

\author[0000-0001-6755-1315]{Joel Leja}
\affiliation{Department of Astronomy \& Astrophysics, The Pennsylvania State University, University Park, PA 16802, USA}
\affiliation{Institute for Computational \& Data Sciences, The Pennsylvania State University, University Park, PA 16802, USA}
\affiliation{Institute for Gravitation and the Cosmos, The Pennsylvania State University, University Park, PA 16802, USA}

\author[0000-0001-9002-3502]{Danilo Marchesini}
\affiliation{Department of Physics \& Astronomy, Tufts University, MA 02155, USA}

\author[0000-0001-8367-6265]{Tim B. Miller}
\affiliation{Center for Interdisciplinary Exploration and Research in Astrophysics (CIERA), Northwestern University, 1800 Sherman Ave, Evanston IL 60201, USA}

\author[0000-0001-7300-9450]{Ikki Mitsuhashi}
\affiliation{Department for Astrophysical and Planetary Science, University of Colorado, Boulder, CO 80309, USA}

\author[0000-0003-2804-0648 ]{Themiya Nanayakkara}
\affiliation{Centre for Astrophysics and Supercomputing, Swinburne University of Technology, Melbourne, VIC 3122, Australia}

\author[0000-0002-7524-374X]{Erica J. Nelson}
\affiliation{Department for Astrophysical and Planetary Science, University of Colorado, Boulder, CO 80309, USA}

\author[0000-0002-9651-5716]{Richard Pan}
\affiliation{Department of Physics \& Astronomy, Tufts University, MA 02155, USA}

\author[0000-0002-0108-4176]{Sedona H. Price}
\affiliation{Department of Physics and Astronomy and PITT PACC, University of Pittsburgh, Pittsburgh, PA 15260, USA}

\author[0000-0002-1714-1905]{Katherine A. Suess}
\affiliation{Department for Astrophysical and Planetary Science, University of Colorado, Boulder, CO 80309, USA}

\author[0000-0001-9269-5046]{Bingjie Wang (\begin{CJK*}{UTF8}{gbsn}王冰洁\ignorespacesafterend\end{CJK*})}
\affiliation{Department of Astronomy \& Astrophysics, The Pennsylvania State University, University Park, PA 16802, USA}
\affiliation{Institute for Computational \& Data Sciences, The Pennsylvania State University, University Park, PA 16802, USA}
\affiliation{Institute for Gravitation and the Cosmos, The Pennsylvania State University, University Park, PA 16802, USA}

\begin{abstract}\noindent
Low-mass quiescent galaxies are thought to predominantly reside in overdense regions, as environmental effects are often invoked to explain their shutdown of star formation. These longer-timescale quenching mechanisms - such as interactions with hot gas in the intracluster medium and dynamical encounters with other cluster galaxies - leave imprints on galaxy morphologies, emphasizing the importance of quantifying the structures of low-mass quiescent galaxies in galaxy clusters at $z<0.5$. Using spectrophotometric data from the UNCOVER and MegaScience programs, we present the first measurement of the quiescent size-mass relation between $7<\log(M_\star/M_\odot)<10$ using JWST imaging, based on a sample of 1531 galaxies in the $z=0.308$ Abell 2744 galaxy cluster. The resulting size-mass relation has a significantly higher scatter than similar-redshift field samples, despite comparable best-fit relations in both the dwarf and intermediate-mass regimes. Both ``progenitor bias,'' where larger, diskier low-mass galaxies enter the cluster at later epochs, and a general expansion of galaxy structure from dynamical interactions could be at play. This evolutionary framework is further supported by the tentative evidence that older low-mass quiescent galaxies in the cluster are more spheroidal. The star-formation histories derived for our cluster sample imply their formation and quenching occurs relatively late, at $z<1.5$. In this scenario, the progenitor population would have disky axis-ratio distributions at cosmic noon, in agreement with recent observations. While this leaves ample time for dynamical interactions to maintain quiescence and drive the observed subsequent morphological evolution post-quenching, the data disfavors an onset of quenching due to the environment.

\end{abstract}


\section{Introduction}
Understanding the physical mechanisms driving the quenching of star formation in galaxies is a primary area of research in the field of galaxy evolution. Over the past several decades, empirical studies have established many mechanisms that may be responsible for quenching massive galaxies: active galactic nuclei feedback \citep[e.g.,][]{Croton2006,Choi2018}, bulge formation and other morphological changes \citep{Martig2009}, shock heating of circumgalactic gas \citep[e.g.,][]{Dekel2006,Dekel2019}, and/or cosmological starvation \citep{Feldmann2015,Feldmann2016}. However, these processes are all strongly mass dependent and dominate at high masses \citep[``mass quenching,''][]{Baldry2006,YPeng2010,Kawinwanichakij2017}, and thus cannot explain the existence of the increasing numbers of low-mass quiescent galaxies discovered with the James Webb Space Telescope \citep[JWST, e.g.,][]{Gelli2023,Marchesini2023,Cutler2024,Hamadouche2024}. 

The commonly invoked explanation has long been that low-mass galaxies are quenched by environmental effects \citep{Einasto1974,Boselli2006,Boselli2014,Haines2007,Geha2012}, including interactions with other galaxies (e.g., dynamical stirring/harassment) or with the intracluster medium \citep[ICM, e.g., ram pressure stripping or starvation,][]{Gunn1972}. In particular, since short-timescale interactions with the intracluster medium (ram-pressure stripping) should not result in significant structural changes, recent morphological studies that find a flatter slope of the quiescent size-mass relation at low stellar masses \citep[relative to the massive end, e.g.,][]{Dutton2011,Mosleh2013,Lange2015, Whitaker2017,Nedkova2021,Cutler2022,Cutler2024,Yoon2023, Hamadouche2024} are interpreted as direct empirical evidence supporting this paradigm. These studies cover a wide range of redshifts, masses, and environments, suggesting this flatter slope, similar to that of the star-forming size-mass relation, is a physical feature of the quiescent galaxy population and not driven by systematics. Similarly, low-mass quiescent galaxies at cosmic noon have also been observed with disk-like structures and lower S\'ersic indices with remarkably young (light-weighted) stellar populations overall at cosmic noon \citep[e.g.,][]{Cutler2024,Hamadouche2024}. 

If the onset of environmental quenching occurs at cosmic noon \citep[e.g.,][]{Ji2018,Cutler2024}, these low-mass galaxies should have significantly different morphologies by $z\sim0$ due to the effects of residing in overdense environments. Indeed, studies that examine the size and structure of local dwarf satellites ($\log(M_\star/M_\odot)<8.5$) find they are predominantly spheroidal (based on axis-ratio distributions) with a steeper size-mass relation \citep{Eigenthaler2018}. However, other studies of dense environments at $0.01<z<0.04$ find the $\log(M_\star/M_\odot)>9.4$ quiescent size-mass relation shows similar morphological properties to cosmic noon samples \citep{Yoon2023}. Several other studies of quiescent galaxy size-mass relations in cluster environments are similarly limited to the most massive end \citep[e.g.][]{Kuchner2017,Afanasiev2023,Yang2024}.

Studies of low-mass quiescent galaxies at $z<0.5$ to date are heterogeneous in nature owing to various observational limitations. In general, surface brightness sensitivity has limited measurements of the size-mass relation in clusters to $\log(M_\star/M_\odot)\gtrsim9.5$, while studies of dwarf galaxies at $\log(M_\star/M_\odot)<8$ are very nearby and utilize specialized, low-surface-brightness searches. \cite{Nedkova2021} use deep Hubble Space Telscope (HST) data from the Hubble Frontier Fields \citep[HFF,][]{Lotz2017} to measure galaxy sizes down to $\log(M_\star/M_\odot)\sim7$, but this sample is not specific to cluster galaxies as it also includes CANDELS \citep{Grogin2011,Koekemoer2011} field galaxies and sources at lower redshifts than the clusters. As such, size-mass studies specific to overdense environments only sparsely cover masses between $8\lesssim\log(M_\star/M_\odot)\lesssim9.5$ \citep[see, e.g., Fig. 8 in][]{Eigenthaler2018}, though the advent of JWST imaging has recently opened this parameter space. This is a critical stellar mass range thought to contain the low-mass peak of the quiescent stellar mass function at $z\gtrsim0.5$ \citep[e.g.,][]{Ilbert2013,Santini2022,Weaver2023,Hamadouche2024}, below which quiescent galaxies potentially become more rare.

The star-formation histories (SFHs) of low-mass, low-redshift cluster galaxies offer crucial insights as to whether or not the now observationally-accessible low-mass quiescent population at the peak epoch of star formation \citep[e.g.,][]{Cutler2024} are direct progenitors. In particular, non-parametric SFHs \citep[e.g.,][]{Leja2019} can recover the potentially bursty SFHs many low-mass galaxies are theorized to have \citep[e.g.,][]{Tacchella2016,Angles-Alcazar2017,Faucher-Giguere2018,Ma2018,Looser2023,Looser2024,Sun2023,Cenci2024,Dome2024,Gelli2025,Trussler2025,Mintz2025}, resulting in more accurate median stellar ages and formation redshifts. SFHs are most often inferred from detailed spectral energy distribution (SED) fitting, however, stellar population modeling from broadband photometry alone is unable to constrain bursts of star formation on 100 Myr to 1 Gyr timescales (e.g., \citealt{Suess2022}; \citealt{Wang2024}; G. Khullar et al. in prep.), meaning that accurate measurements of a galaxy's SFH requires a significant amount of high-spectral-resolution panchromatic data (i.e., medium bands and/or spectroscopy).

In this paper, we perform a comprehensive analysis of the size, structure, and formation histories of low-mass quiescent galaxies in the $z\sim0.3$ cluster environment, and compare to recent observations of cosmic noon overdensities to investigate the formation and evolution of low-mass quiescent galaxies in the $\sim10$ Gyr since $z\sim2$. Specifically, we leverage the extensive spectrophotometric suite of JWST data associated with the Ultradeep NIRSpec and NIRCam ObserVations before the Epoch of Reionization (UNCOVER) Cycle 1 Treasury program \citep{Bezanson2024}. UNCOVER observations cover the Abell 2744 lensing cluster, making it an ideal survey to study low-mass quiescent galaxies in rich overdensities. Aside from having extremely deep NIRCam broadband photometry \citep{Weaver2024}, UNCOVER also has NIRSpec prism spectroscopy \citep{Price2024} and associated NIRCam medium band photometry from the Medium Bands, Mega Science Survey \citep{Suess2024}. Medium bands are crucial in both constraining the shape of the galaxy's SED and providing additional short-wavelength coverage, while the prism spectroscopy places strong constraints on the inferred parameters from SED modeling via joint spectrophotometric fitting \citep[e.g.,][]{prospector2021,Akhshik2023}.

We structure this paper as follows. The UNCOVER and MegaScience data and cataloging, as well as our sample selections, are detailed in Section \ref{sec:sample}. Our morphological fitting methods and stellar population modeling are described in Section \ref{sec:methods}. In Section \ref{sec:results}, we show our Abell 2744 quiescent size-mass relation and the results of our spectrophotmetric SED fitting. Using these results, we contextualize our sample in the overall growth and evolution of low-mass quiescent galaxies since cosmic noon in Section \ref{sec:discuss}. The study is summarized in Section \ref{sec:summary}. Magnitudes are in the AB system. We assume a \cite{Chabrier2003} initial mass function and a WMAP9 cosmology \citep{Hinshaw2013}: $H_0=69.32~{\rm km~s^{-1}~Mpc^{-1}}$, $\Omega_M=0.2865$, and $\Omega_\Lambda=0.7135$.

\section{Data and Sample Selection}\label{sec:sample}
\subsection{Imaging and Catalogs}
This study utilizes observations of the Abell 2744 cluster from the UNCOVER survey \citep{Bezanson2024,Weaver2024} with the extended photometric coverage of the Medium Bands, Mega Science Survey \citep[hereafter MegaScience,][]{Suess2024}. The resulting dataset contains photometry from JWST/NIRCam in the F070W, F090W, F115W, F150W, F200W, F277W, F356W, and F444W broadband filters, as well as the F140M, F162M, F182M, F210M, F250M, F300M, F335M, F360M, F410M, F430M, F460M, and F480M medium-band filters. We also include archival HST data from the HFF \citep{Lotz2017} and BUFFALO \citep{Steinhardt2020} surveys in the ACS F435W, F606W, and F814W and WFC3 F105W, F125W, F140W, and F160W broadband filters. The data presented in this article were obtained from the Mikulski Archive for Space Telescopes (MAST) at the Space Telescope Science Institute; the specific observations analyzed for UNCOVER and MegaScience can be accessed at \dataset[DOI: 10.17909/16ez-hg64]{https://doi.org/10.17909/16ez-hg64} and \dataset[DOI: 10.17909/8k5c-xr27]{https://doi.org/10.17909/8k5c-xr27}. UNCOVER and MegaScience imaging is from the \grizli{}-reduced v7\footnote{\url{https://dawn-cph.github.io/dja/imaging/v7/}} mosaics \citep{grizli} and rescaled to a 40 mas pixel scale in all filters. 

At the redshift of the Abell 2744 galaxy cluster ($z=0.308$), the JWST filters sample rest-frame $0.5-3.4~{\rm\mu m}$. Thus, galaxies that are red in the rest-optical within Abell 2744 appear relatively blue in our filter set. In order to ensure an effective detection of all cluster galaxies, we build our own short-wavelength (SW) detected catalogs using the \textsc{Aperpy}\footnote{\url{https://github.com/astrowhit/aperpy}} aperture photometry code. All \textsc{Aperpy} settings are identical to \cite{Weaver2024}, except the detection image, for which we use a F070W-F090W-F115W selection function (at \ang[angle-symbol-over-decimal]{ ; ;0.04} pixel$^{-1}$) and detect on imaging with added median filtering. In order to avoid a changing selection function across the image, we only consider pixels where all three filters have data. This SW detection better selects galaxies at the cluster redshift ($z=0.308$), while the median filtering assists with the detection of faint sources in areas that still have significant intracluster light (ICL) or residuals after the model-subtraction of bright cluster galaxies (bCGs; see Z. Webb et al. in prep). Using this SW detection, we find 165 (10\%) additional sources over the fiducial UNCOVER/MegaScience F277W-F356W-F444W detection from \cite{Weaver2024}. Photometric measurements, however, are made on F444W PSF-matched imaging with the bCGs, ICL, and sky background subtracted \citep[see][]{Weaver2024}.

Initial redshifts and stellar population parameters for the SW detected catalog are determined with \eazy{} \citep{Brammer2008} using two template sets: the default \textsc{fsps\_full} and the \textsc{sfhz\_blue\_agn} set. \textsc{sfhz\_blue\_agn} makes use of several additions to the \textsc{fsps\_full} templates, including redshift-dependent priors on galaxy age and SFH, empirical spectra from a $z=8.5$, emission-line galaxy \citep{Carnall2023}, and an AGN torus template \citep{Killi2023}. The optional magnitude and $\beta$-slope (i.e., UV slope) priors are disabled in the fits. The typical photometric redshift uncertainty at $z\sim0.3$ is $\delta z=0.06$. This uncertainty is irrespective of galaxy selection, but only considers galaxies brighter than 27 ABmag in F200W (the minimum brightness for our sample selection, see Section \ref{sec:selection}). 

\begin{figure}
    \centering
    \includegraphics[width=0.98\linewidth]{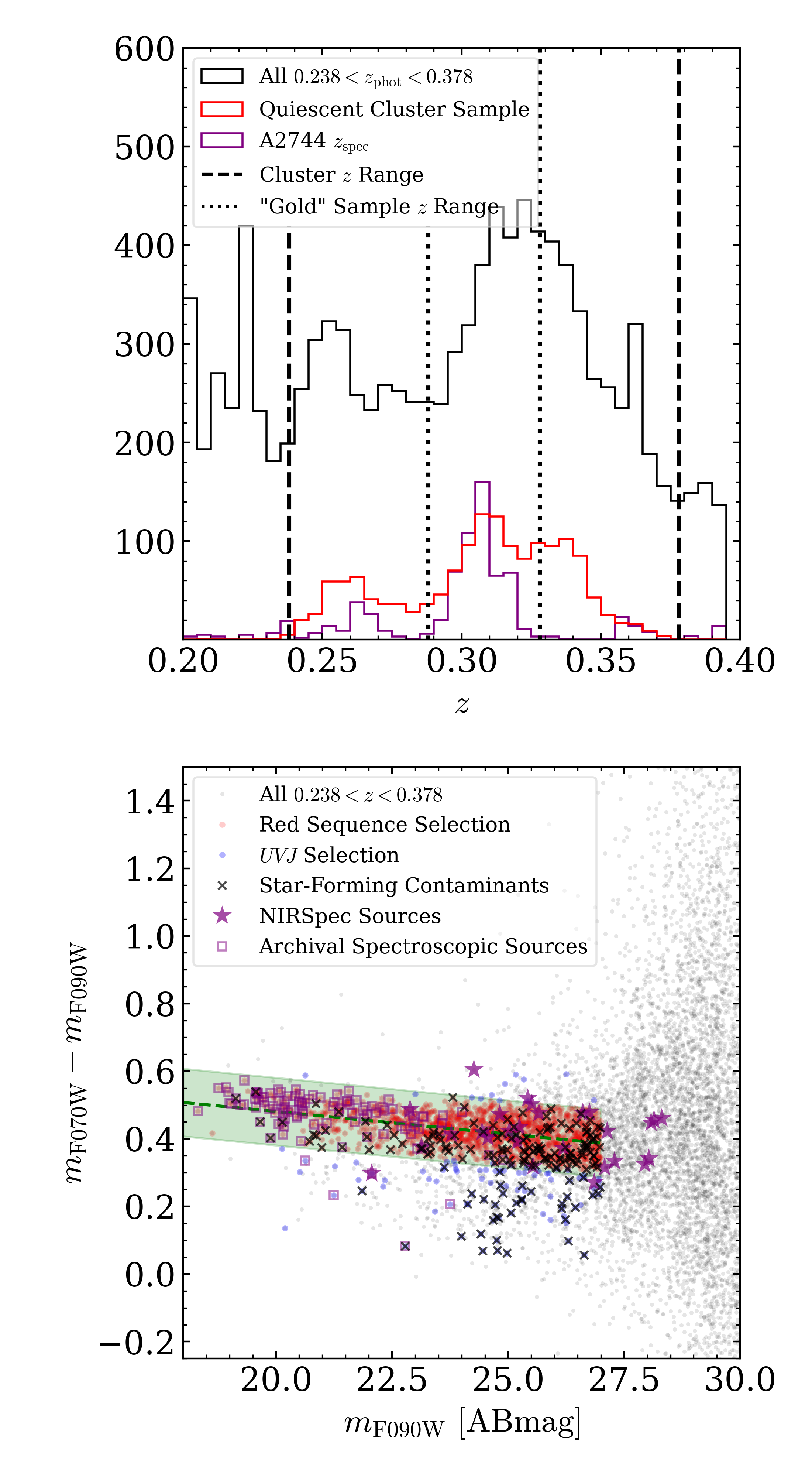}
    \caption{The selected sample of low-mass cluster galaxies from the UNCOVER/MegaScience SW-selected catalogs. Top: The redshift distributions for all $0.238<z_{\rm phot}<0.378$ galaxies and the sample of cluster galaxies selected via red sequence fitting are shown in black and red, respectively. The purple histogram shows the distribution of all confirmed $0.238<z_{\rm spec}<0.378$ spectroscopic redshifts in Abell 2744 obtained from \cite{Kokorev2022}. The dashed black lines highlight the redshift range used to select cluster galaxies while dotted black lines indicate the more constrained redshift range used to select a ``gold'' sample of cluster members. Bottom: The positions of all sources at $0.238<z<0.378$ in the F070W$-$F090W color-magnitude diagram are shown with black points and the red-sequence selection of quiescent cluster members is highlighted in red. Additional quiescent cluster members selected with a $UVJ$ selection are indicated with blue points. The green dashed line indicates the best fit line to the red sequence with the shaded region highlighting the $2\sigma$ scatter (roughly 0.1 ABmag). Purple stars and x's represent sources with corresponding NIRCam PRISM spectroscopy and other archival spectroscopic redshifts, respectively.}
    \label{fig:sample}
\end{figure}

\subsection{Sample Selection}\label{sec:selection}
Classically, quiescent cluster members are selected from the red-sequence of the color-magnitude diagram \citep[CMD, e.g.,][]{Visvanathan1977,Bower1992,Bower1992b,Stott2009,Valentinuzzi2011}, including in several studies of Abell 2744 \citep[e.g.,][]{Lee2016,Furtak2023,Harvey2024,Price2024}, compared to color-color selections like $UVJ$. The use of a CMD-based selection is further justified by the fact that NIRCam filters do not offer sufficient coverage of the rest-frame $U$ band at the redshift of Abell 2744 ($z=0.308$), which prevents a robust selection of quiescent galaxies using $UVJ$ cuts. Thus, we select quiescent cluster members using a combined F070W$-$F090W CMD and redshift selection. The F070W$-$F090W color covers the rest-optical ($\sim5350-6880$ \AA{}), which can be used to distinguish quiescent cluster galaxies from star-forming ones \citep[e.g. Fig. 8,][]{Price2024}. The dashed lines in top panel of Figure \ref{fig:sample} show our redshift selection; we select all \ninit{} sources with \eazy{} photometric redshifts between $0.238<z<0.378$ ($\Delta z<0.07$ or $\Delta D_C\lesssim264$ Mpc) using the \textsc{fsps\_full} templates, excluding any sources marked as artifacts or with bad photometry (\texttt{USE\_PHOT}$=0$) in the photometric catalog. This redshift range, which is slightly larger than the typical photometric redshift uncertainty ($\delta z=0.06$), covers the other peaks of the redshift distribution around Abell 2744 (see Fig. \ref{fig:sample}, top). The \textsc{sfhz\_blue\_agn} template is used to remove high-redshift interlopers due to its better performance at $z\gtrsim1$. These templates, however, have significant biases at $z\lesssim1$, where the \textsc{fsps\_full} templates perform better \citep[see Section 5.3 in][]{Weaver2024} and are thus used in the actual redshift selection. \nphotrem{} galaxies with $z_{\rm phot}>2$ in the \textsc{sfhz\_blue\_agn} templates are removed from the initial redshift selection (\nphotremfrac{}\%). Several sources have archival spectroscopic redshifts (\citealt{Owers2011}; \citealt{Richard2014}; \citealt{Mahler2018}; GLASS \footnote{\url{https://archive.stsci.edu/prepds/glass/}}) obtained from the compilation of \cite{Kokorev2022}. We include any source with $0.238<z_{\rm spec}<0.378$ in the initial redshift selection, regardless of its photometric redshift (purple squares in Fig. \ref{fig:sample}, bottom). This adds an additional \nspeconly{} sources to the photometric redshift selection.

The total redshift distribution of selected sources is shown in the top panel of Figure \ref{fig:sample} for all $0.238<z<0.378$ galaxies in black. The purple histogram shows the distribution of archival spectroscopic redshifts in Abell 2744 from \cite{Kokorev2022}. The spectroscopic redshift distribution indicates that there are other structures in the UNCOVER field-of-view at $z\sim0.265$ and $z\sim0.36$, which we also see in the photometric redshift distributions. Moreover, the $z_{\rm phot}$ distributions are also broader than the $z_{\rm spec}$ distribution, with more sources at slightly higher redshifts. We include all of these sources in our redshift sample in order to maximize sample size, however we also create a ``gold'' sample (dotted lines in Fig, \ref{fig:sample}, top) of \ninitgold{} sources whose photometric or spectroscopic redshift is encapsulated by the central peak at $0.288<z<0.328$ in the spectroscopic redshift distribution (purple histogram in Fig. \ref{fig:sample}, top). While this $\Delta z=0.02$ ($\Delta D_C\lesssim74$ Mpc) still covers larger physical volumes of space than the cluster occupies, the widths of both the $z_{\rm phot}$ and $z_{\rm spec}$ distributions are likely inflated by uncertainties in fitting redshift solutions to the relatively featureless rest-frame NIR spectra of quiescent galaxies.

The red sequence of the CMD is then fit for bright sources ($m_{\rm F090W}<27$) using a linear least-square fitting with 2$\sigma$ outlier removal (equivalent to $\Delta(m_{\rm F070W}-m_{\rm F090W})=0.1$ ABmag). The resulting linear relation is shown with a green line in Figure \ref{fig:sample}, bottom, with the shaded region indicating 0.1 ABmag around the linear fit to the red sequence. The resulting CMD-selected sample (red points in Fig. \ref{fig:sample}) contains \ncmd{} total galaxies and corresponds to a gold sample of \ncmdgold{} galaxies. We limit the selection to sources brighter than $m_{\rm F090W}<27$ to minimize contamination from faint background sources; if we consider a more pure selection of sources brighter than $m_{\rm F090W}<26$ we find no significant change in our results. 

Part of our selection area does contain rest-$U$ photometry from HST/ACS F435W imaging in the HFF \citep[]{Lotz2017,Shipley2018} and BUFFALO \citep{Steinhardt2020} surveys. Thus, in regions of the mosaic with significant ($S/N>3$) F435W coverage, we are able to robustly select quiescent galaxies at $0.238<z<0.378$ using the $UVJ$ cuts from \cite{Cutler2024}. These $UVJ$-selected sources, indicated by blue points in Figure \ref{fig:sample}, are included in the color-selected cluster sample, adding an additional \nuvj{} galaxies (with \nuvjgold{} added to the gold sample).

The color-based selections described above serve as a pre-selection to maximize completeness when selecting a sample of quiescent cluster galaxies. As such, this sample is likely not pure. To ensure a pure sample of quiescent galaxies, the color-selected sample of cluster galaxies is modeled using the \pros{} emulator \citep{Mathews2023} with the \prosa{} model \citep{Leja2017}. Modeling includes 27 photometric filters (where available), including archival HST ACS and WFC3 data and medium- and broad-band photometry from MegaScience and UNCOVER, respectively. Quiescent galaxies are then selected as sources whose average star-formation rate (SFR) over the most recent 100 Myr is $>0.7$ dex below the \cite{Leja2022} star-forming main sequence. This selection removes \nmsrem{} star-forming galaxies (\nmsremfrac{}\% of the color-selected sample), which are indicated by black x's in Figure \ref{fig:sample}.

A subset of UNCOVER sources have JWST/NIRSpec PRISM spectroscopic follow-up observations, which covers an approximate wavelength range of $0.6-5.3\mu$m at a spectral resolution of $R\sim100$. We cross-match all sources with a redshift quality flag $\geq2$ from the UNCOVER DR4 spectroscopic catalog \citep[see][]{Price2024} to our SW-selected catalog within a \ang[angle-symbol-over-decimal]{ ; ;0.24} radius threshold in order to determine which sources in our catalog have associated spectroscopic measurements. We then select all associated NIRSpec sources at $0.2<z_{\rm spec}<0.378$ and verify their spectral shape by eye. The publicly-released DR4 PRISM spectra are also used in our spectrophotometric modeling (see Section \ref{sec:sed}). \nnirspectot{} NIRSpec sources with reasonable spectra and redshifts are found, shown as purple stars in Figure \ref{fig:sample}, including \nnirspecadd{} additional sources that were missed via the photometric selection. The final sample of quiescent cluster galaxies contains \ncluster{} (including \nclustergold{} gold sources) and is indicated in the top panel of Figure \ref{fig:sample} with a red histogram.

\section{Analysis}\label{sec:methods}
\subsection{Structural Modeling}
Galaxies are modeled using a single 2D S\'ersic profile with \galfit{} \citep{Peng2002,Peng2010} following the procedure of \cite{Cutler2024}, which is based on the methods of \cite{vanderWel2012}. The parameter constraints on S\'ersic index are slightly different to avoid overestimates due to ICL ($0.2<n<8$, as opposed to $0.2<n<10$ in \citealt{Cutler2024}); other constraints are identical ($\pm3$ mag from the photometric catalog magnitude, half-light radius within $0.01<R_e<400$ pixels, and axis ratio between $0.0001<q<1$). The modeling is performed on the bCG-subtracted F200W imaging at native resolution, scaled to an \ang[angle-symbol-over-decimal]{ ; ;0.04} ${\rm pix^{-1}}$ pixel scale. F200W is chosen as the SED of these $z=0.308$ galaxies peaks at $\sim2~\mu$m. In addition to the \nflag{} galaxies removed due to sparse coverage or crowding from nearby sources (\nflagfrac{}\% of the initial sample), as well as the \ngalfail{} removed due to bad \galfit{} models \citep[\ngalfailfrac{}\%, see][]{Cutler2024}, we remove any source with an extremely low ($<0.1~M_\odot~{\rm pc}^{-2}$) effective surface-mass density ($\Sigma_{e,M_\star}=M_\star/(2\pi R_{\rm e,SMA}^2)$), as this is indicative of a bad fit. This cut removes \nlowsigma{} galaxies (\nlowsigmafrac{}\% of the initial sample).

We also apply a cut on the residual flux fraction (RFF), similar to \cite{Ormerod2024}, which is determined to be RFF$>0.05$ through visual inspection of model residuals. \nrff{} additional galaxies (\nrfffrac{}\% of the initial sample) are excluded due to high RFF. While this moderately-high rejection fraction suggests S\'ersic models are not the most suitable approach, we believe it is a consequence of the data at hand, with most sources at $\log(M_\star/M_\odot)<9$ failing primarily due to contamination from nearby sources or other ICL. The morphologies of most sources at $\log(M_\star/M_\odot)>10$ are indeed too complex to be effectively modeled by a single S\'ersic profile at these redshifts, leading to high RFFs and a significantly higher rejection rate (\hmfrac{}\% compared to \lmfrac{}\% at $\log(M_\star/M_\odot)<9$). As such we impose a general $10^{10}~M_\odot$ upper limit on stellar mass (following the RFF and $\Sigma_{e,M_\star}$ cuts) to avoid significantly biasing population statistics at these masses and because our primary science goals are focused on lower stellar masses. This removes \nmasslim{} additional galaxies (\nmasslimfrac{}\%); in total \nhimass{} sources are removed at $\log(M_\star/M_\odot)>10$. The resulting final morphological sample contains \nfinal{} galaxies, \nfinalgold{} of which are from the gold sample. The results shown and discussed in the paper are with respect to the full sample. We find no significant changes when we only consider sources in the gold sample.

Though a model-based measurement of $R_e$ might be less efficient at handling complex morphologies or contamination from nearby sources, it is much more accurate at accounting for extended light below the noise limit of the imaging, which is likely present in these low-mass, diffuse galaxies. As such, we primarily use the \galfit-based sizes for all \nfinal{} sources left following quality cuts. However, for completeness, we measure the sizes of all sources removed by these cuts via individual growth curves, giving us size measurements for all \ncluster{} galaxies in our cluster sample.

\subsection{SED Modeling}\label{sec:sed}
\nnirspectot{} galaxies in our sample have NIRSpec PRISM spectroscopic observations (see Section \ref{sec:sample}). For these sources, we jointly model the spectrophotometric data using the Bayesian SED fitting code \pros{} \citep{prospector2021}. The SED-fitting procedure closely follows \cite{deGraaff2025}. We utilize the default \pros{} libraries: the Flexible Stellar Population Synthesis models \citep[FSPS,][]{fsps2009,fsps2010a,fsps2010b}, the MILES spectral libraries \citep{miles2006}, and the MIST stellar isochrones \citep{mist2016a,mist2016b}. We also incorporate non-parametric SFH based on the ``continuity'' prior \citep{Leja2019}, which fits for the logarithmic ratio between adjacent age bins and applies Student's-t priors centered on 0 with a width of 0.3 and $\nu=2$ degrees of freedom to these ratios. The SFH is measured over 14 age bins, where the first 100 Myr are divided into bins of width 10, 40, and 50 Myr and the remaining 11 linearly spaced bins (921 Myr widths). Sampling is done using nested sampling from \textsc{dynesty} \citep{Speagle2020}.

Dust attenuation is modeled using the \cite{Kriek2013} curves, which additionally fit for the ``dust index'', an offset to the \cite{Calzetti2000} attenuation slope. We also assume twice as much young star ($t<10$ Myr) attenuation when compared to older stars. Dust emission is set using the \cite{Draine2007} models, where the mass fraction of dust in polycyclic aromatic hydrocarbon and dust in high intensity radiation is fixed at 2 and 1\%, respectively, and the interstellar radiation strength is at the Milky Way value. Stellar metallicity is also a free parameter over the range $0.1Z_\odot<2Z_\odot$.

To account for resolution effects, the model spectra are convolved with the PRISM resolution curve found in the JWST user documentation\footnote{\url{https://jwst-docs.stsci.edu/jwst-near-infrared-spectrograph/nirspec-instrumentation/nirspec-dispersers-and-filters}} with an added multiplication factor of 1.1 \citep[e.g.,][]{Curtis-Lake2023,deGraaff2025}. This factor accounts for the extended median size of these sources ($R_{e,{\rm SMA}}=\ang[angle-symbol-over-decimal]{ ; ;0.24}$). Uncertainty in the NIRSpec flux calibration is treated using \pros{} \textsc{PolySpecModel}: a 12th order polynomial is fit to the input spectrum  during each likelihood call in order to match the observed and model spectra.

\begin{figure*}[h!]
    \centering
    \includegraphics[width=\linewidth]{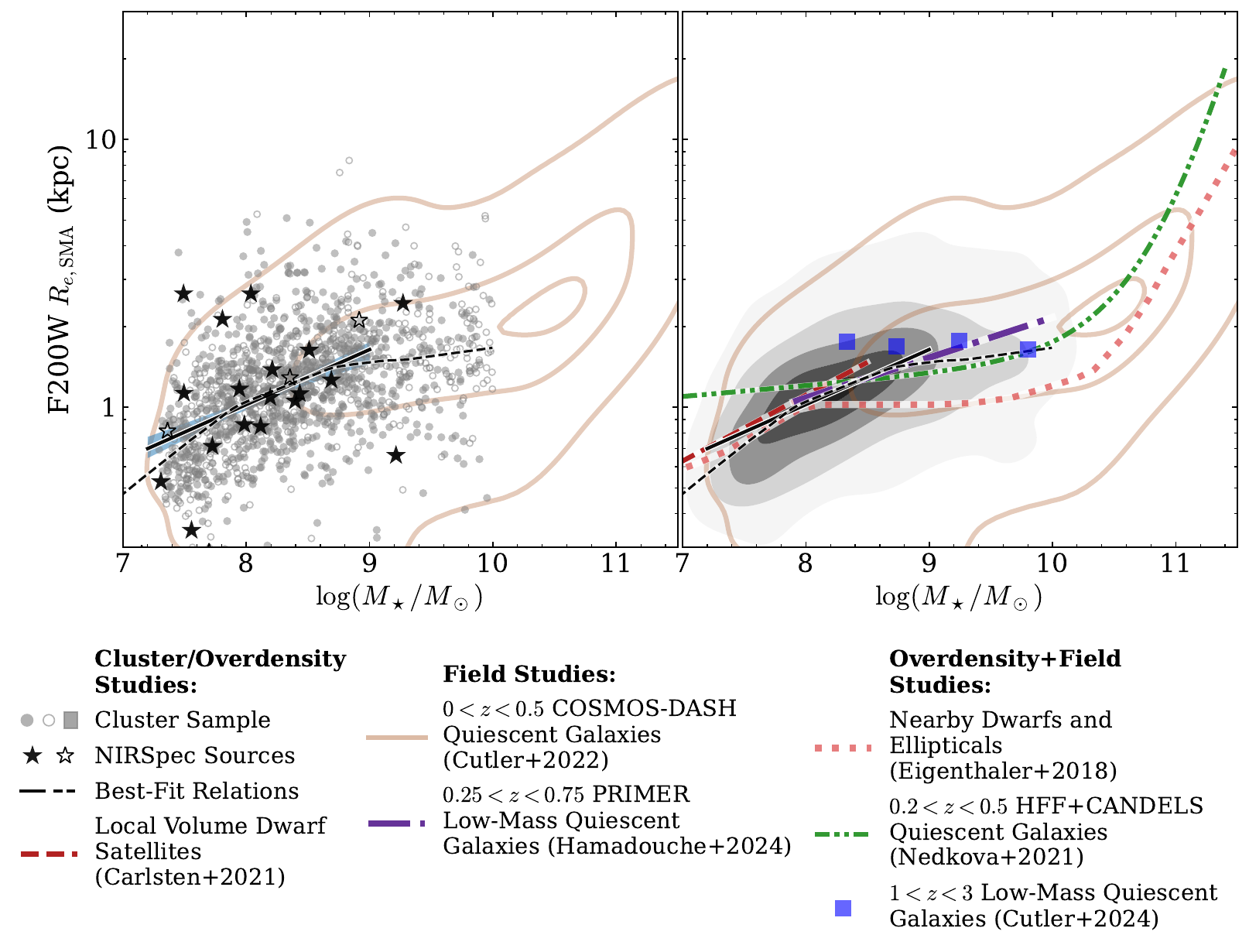}
    \caption{Low-mass quiescent galaxies in Abell 2744 follow a similar size-mass relation to those measured from local dwarf galaxies. However, the larger intrinsic scatter relative to field studies may indicate the effect of dynamical interactions expanding and/or eroding cluster galaxies over time. In the left panel, F200W semi-major axis sizes of low-mass quiescent galaxies in our sample are indicated with gray points. Sources that have corresponding NIRSpec spectroscopy are highlighted with black stars. Open points and stars indicate sources whose sizes are measured via growth curves (due to poor \galfit{} models). In the right panel size-mass relations from the literature are shown: blue squares show the median sizes of $1<z<3$ low-mass quiescent galaxies from \cite{Cutler2024}, a dotted red line shows the fit to a combined sample of nearby dwarfs and massive ellipticals in \cite{Eigenthaler2018}, a dashed maroon line shows the size-mass relation of local dwarf satellites in \cite{Carlsten2021}, a dotted-dashed green line shows the \cite{Nedkova2021} $0.2<z<0.5$ size-mass relation based on HFF and CANDELS, and a dotted-dashed purple line represents the PRIMER $0.25<z<0.75$ low-mass quiescent size-mass relation from \cite{Hamadouche2024}. Gray contours indicate the distribution of all Abell 2744 cluster galaxies in our sample (both solid and open points in the left panel). In both panels, the solid black line shows the linear best fit size-mass relation to \galfit{}-modeled, $\log(M_\star/M_\odot)<9$ quiescent galaxies (Eq. \ref{eqn:msr}), while the dashed black line shows the LOWESS fit to all galaxies in the sample. The shaded cyan region in the left panel shows the $1\sigma$ uncertainty on the linear fit. Brown contours in all panels show the distribution of $0<z<0.5$ quiescent field galaxies from \cite{Cutler2022}. The F200W PSF HWHM is \ang[angle-symbol-over-decimal]{ ; ;0.032}, corresponding to $\sim0.15$ kpc at $z=0.308$.}
    \label{fig:size-mass}
\end{figure*}

\begin{figure*}
    \centering
    \includegraphics[width=\linewidth]{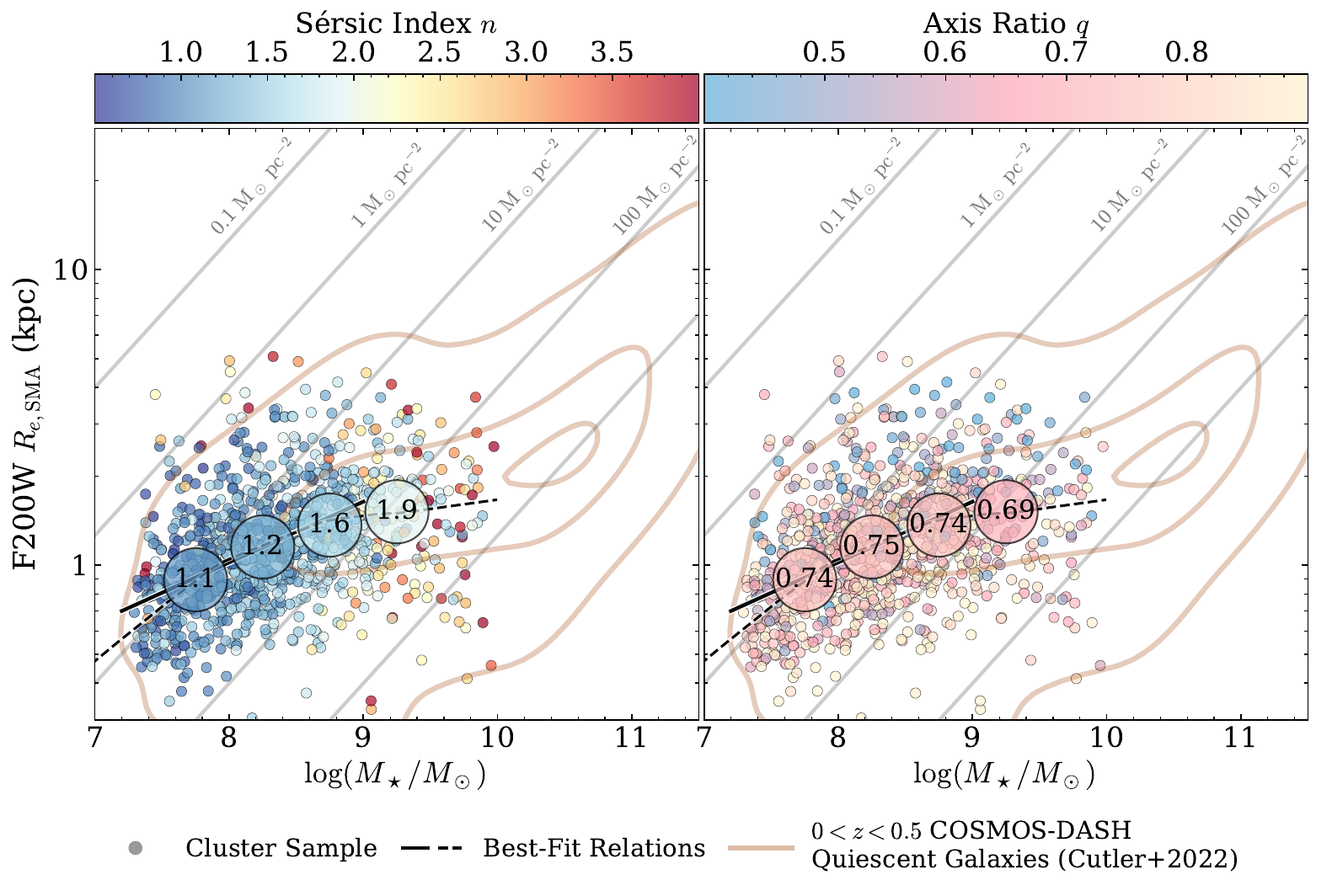}
    \caption{Low-mass quiescent galaxies in Abell 2744 have exponential-disk-like light profiles, though their larger median axis ratios do not indicate an overall disky population. The sample of low-mass quiescent galaxies are colored by their F200W S\'ersic index (left) and axis ratio (right), with large circles indicating the median size in 0.5 dex mass bins. Large numbers show the median S\'ersic index and median axis ratio of the bins, respectively. Solid and dashed black lines indicate the best-fit size-mass relations shown in Figure \ref{fig:size-mass}; brown contours indicate the $0<z<0.5$ quiescent field galaxies from \cite{Cutler2022}. Lines of constant effective surface-mass density ($\Sigma_{\rm e,M_\star}$) are shown in gray, ranging from $0.1<\Sigma_{\rm e,M_\star}<1000~{\rm M_\odot~pc^{-2}}$.}
    \label{fig:struc}
\end{figure*}
\section{Results}\label{sec:results}

\subsection{Low-Mass Quiescent Galaxy Structure}\label{sec:structure}
\subsubsection{The Size-Mass Relation}\label{sec:size-mass}
Several studies have presented quiescent size-mass relations in cluster environments \citep{Kuchner2017,Afanasiev2023,Yang2024}, though these studies are limited to $\log(M_\star/M_\odot)\gtrsim9.5$. Our analysis obtains cluster galaxy sizes at $7<\log(M_\star/M_\odot)<10$, filling in this gap in the literature. In the left panel of Figure \ref{fig:size-mass}, we show the sizes of $\log(M_\star/M_\odot)<10$ quiescent galaxies in Abell 2744 (black points). Sources with no \galfit{} results are also shown with open circles and stars; sizes for these galaxies are estimated from individual growth curves. In the right panel, gray contours indicate the full sample (including growth-curve-based sizes). Following \cite{Hamadouche2024}, we fit the low-mass quiescent size-mass relation {(to all galaxies with \galfit{} measurements)} using a single power law of the form 
\begin{align}\label{eqn:msr}
    \log\left(\frac{R_e}{{\rm kpc}}\right)=\alpha\times\log\left(\frac{M_\star}{5\times10^{10}M_\odot}\right)+\log(A),
\end{align}
finding best-fit values of $\alpha=\alphaval{}\pm\alphaerr{}$ and $\log(A)=\logaval{}\pm\logaerr{}$. 

The right panel of Figure \ref{fig:size-mass} shows that our best-fit relation to the $\log(M_\star/M_\odot)<9$ quiescent galaxies (solid black line) is comparable to the $\log(M_\star/M_\odot)<8$ size-mass relation for dwarf galaxies from the Fornax cluster \citep[dotted red line,][]{Eigenthaler2018} and the Local Volume \citep[dashed maroon line,][]{Carlsten2021}. We also measure the full, $\log(M_\star/M_\odot)<10$ size-mass relation (including sources with failed \galfit{} fits) using a locally-weighted scatterplot smoothing \citep[LOWESS, e.g.,][]{Cleveland1981}, following \cite{Eigenthaler2018}, shown as a dashed black line in Figure \ref{fig:size-mass}. The LOWESS size-mass relation agrees with both the dwarf galaxy relations described above, as well as the intermediate-mass relation from HFF and CANDELS \citep{Nedkova2021} and the size-mass contours of quiescent field galaxies from COSMOS-DASH \citep{Cutler2022}, though it is notably larger than the intermediate-mass relation of \cite{Eigenthaler2018}. However, we measure a model scatter of $\sigma_{\log R_e}=\logsreval{}\pm\logsreerr{}$, a factor of two larger than the $\sim0.1$ dex size scatter in $z<0.5$ quiescent field galaxies \citep{vanderWel2014,Kawinwanichakij2021,Hamadouche2024}. The large scatter we find is not primarily driven by projection effects. If we instead consider circularized sizes ($R_{\rm e,circ}=R_{\rm e,SMA}\times\sqrt{q}$), we find a comparable scatter of $\sigma_{\log R_e}=\logsrevalcirc{}\pm\logsreerrcirc{}$.

\begin{figure*}[ht!]
    \centering
    \includegraphics[width=\linewidth]{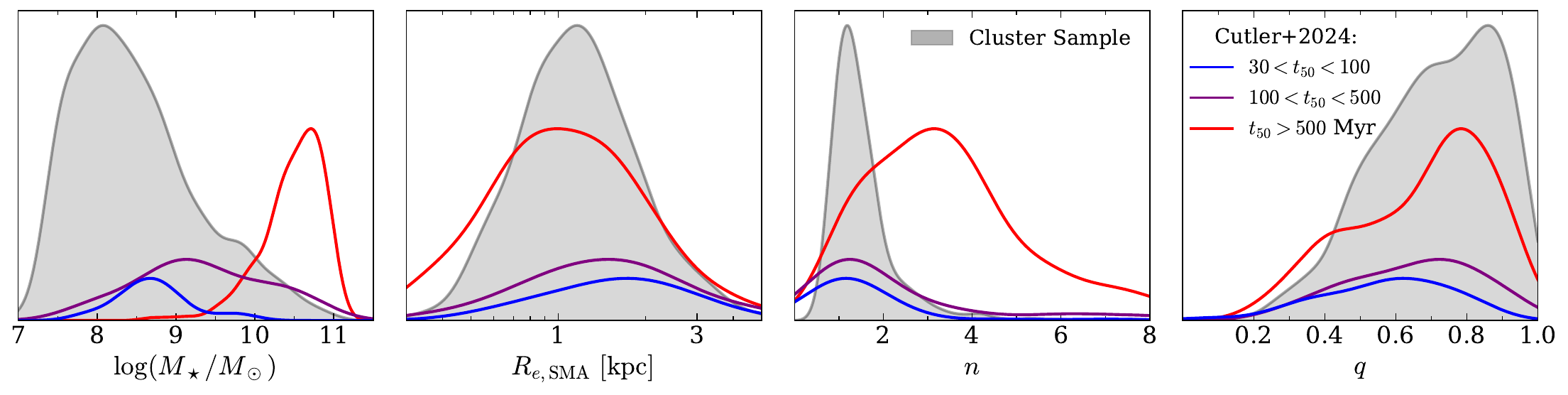}
    \caption{Low-mass quiescent galaxies in Abell 2744 have similar axis ratio distributions to old, massive quiescent galaxies at $1<z<3$. Distributions of mass, size, S\'ersic index, and axis ratio for low-mass quiescent galaxies in Abell 2744 (gray shaded histogram) and cosmic noon quiescent galaxies from \cite{Cutler2024}. The cosmic noon sample is separated by median stellar age: $30<t_{50}<100$ in blue, $100<t_{50}<500$ in purple, and $t_{50}>500$ Myr in red.}
    \label{fig:strucparams}
\end{figure*}

\subsubsection{Axis-Ratio Distributions} \label{sec:axisratio}
The left panel of Figure \ref{fig:struc} indicates that Abell 2744 galaxies predominantly have exponential disk light profiles ($n<2$), similar to \cite{Cutler2024} and \cite{Hamadouche2024}. However, the median axis ratios (Fig. \ref{fig:struc}, right) are slightly larger than the cosmic noon sample from \cite{Cutler2024}: $\sim0.73\pm0.13$ compared to $\sim0.58\pm0.10$. The larger axis ratios suggest that the low-mass quiescent populations has more spheroidal morphologies, potentially driven by billions of years worth of dynamical interactions in the cluster environment. Figure \ref{fig:strucparams} shows the distribution of structural parameters in both Abell 2744 low-mass quiescent galaxies (gray shaded histograms) and $1<z<3$ quiescent galaxies from \cite{Cutler2024} (colored histograms). In terms of mass, size, and S\'ersic index, our sample is most similar to younger ($t_{50}<500$ Myr), low-mass quiescent galaxies at cosmic noon (blue and purple curves). Conversely, we find that the distribution of axis ratios strongly peaks at $q\sim0.7-0.8$; this behavior is similar to old ($t_{50}>500$ Myr), massive quiescent galaxies at $1<z<3$ (red curve), whereas low-mass quiescent galaxies exhibit broad, flat distributions.

Notably, however, only a fraction of the $1<z<3$ sample reside in an overdensity, meaning the full \cite{Cutler2024} sample is not necessarily a direct comparison to our Abell 2744 galaxies. Recently, the JWST Cycle 2 grism survey All the Little Things \citep[ALT,][]{Naidu2024} has provided robust grism redshifts out to $z\sim9$ across the Abell 2744 lensing field and in doing so identified several overdensities at high redshifts, including a significant $z\sim2.5$ overdensity \citep{Naidu2024,Whitaker2025} that is studied in-depth by \cite{Pan2025}. We find that a significant fraction (73\%) of the $7<\log(M_\star/M_\odot)<10$ quiescent galaxies from \cite{Cutler2024} at $z>2$ are within $\Delta z=0.2$ (roughly equivalent to the typical photometric redshift uncertainty of this sample) of $z=2.5$, indicating they are likely a part of this overdensity. Combined with the structural and stellar population analysis, these results support the idea that, taking into account the redshift distribution of the \cite{Cutler2024} sample (with 73\% at $z>2$), at least a subset (16\%) of the overall \cite{Cutler2024} sample may occupy a cluster environment at lower redshifts. Sources in the overdensity have a median $q=0.65\pm0.14$ and $n=1.1\pm0.2$, suggesting that low-mass quiescent galaxies in the $z\sim2.5$ overdensity are comparable structurally to those in the field at $1<z<3$.

\subsection{Low-Mass Quiescent Galaxy Formation Histories} \label{sec:form}
At $1<z<3$, a majority of low-mass quiescent galaxies have $t_{50}<500$ Myr \citep[e.g., Fig. 6 in][but see caveat in Sec. \ref{sec:discuss}]{Cutler2024}. While this age distribution may be due to selection effects (older, faint red sources are harder to detect), it could also be physically motivated. Several studies of the quiescent galaxy stellar mass function find that low-mass quiescent galaxies only start to appear at $z\sim2.5$ \citep[e.g.,][]{Santini2022,Hamadouche2024}, suggesting that the lack of old, low-mass quiescent galaxies is simply due to the fact that they have only just begun to quench at cosmic noon. Environmental quenching processes (both longer timescale gas heating and more rapid gas removal) are largely dependent upon the existence of a hot ICM in the overdensity, which takes time to be established. Thus, the existence of low-mass quiescent galaxies prior to that of (proto-)clusters with hot ICM would rule out these quenching mechanisms, at least in the case of onset (as opposed to physical mechanisms that maintain quiescence). In general, there are very few literature studies focused on $\log(M_\star/M_\odot)<9$ quiescent galaxies in $z>1$ (proto-)clusters, as previous studies are typically sensitive to quiescent galaxies down to $\log(M_\star/M_\odot)\sim10$ \citep[e.g.,][]{Papovich2012,Cooke2016,Lee-Brown2017,vanderBurg2020,Leste2024}. However, JWST-based studies \citep[e.g.,][]{Cutler2024,Pan2025} are able to detect quiescent galaxies in cosmic noon overdensities down to $\log(M_\star/M_\odot)\sim8$. As such, comparisons to these works are key in this analysis, despite the large gap in cosmological time between $z=0.308$ and $z\sim2.5$. 

\begin{figure}
    \centering
    \includegraphics[width=\linewidth]{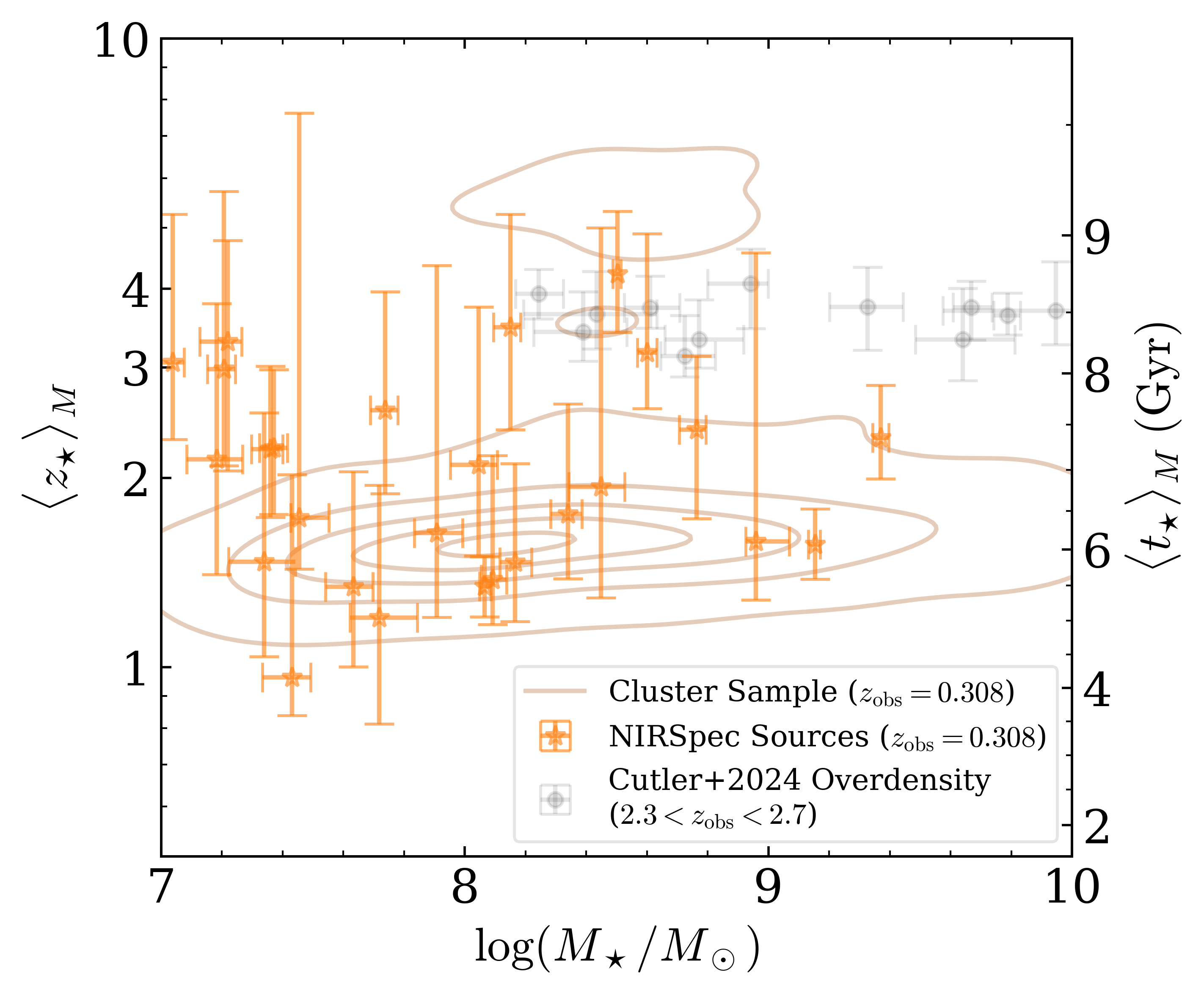}
    \caption{Low-mass quiescent galaxies in Abell 2744 form at lower redshifts compared to those in a $z\sim2.5$ overdensity. Orange stars show the mass-weighted formation redshifts ($\langle z_\star\rangle_M$) of low-mass quiescent cluster galaxies with associated NIRSpec observations while black points show low-mass quiescent galaxies in the $z\sim2.5$ overdensity behind Abell 2744. Brown contours show the underlying sample of all low-mass quiescent galaxies in Abell 2744. The right axis indicates the corresponding mass-weighted age ($\langle t_\star\rangle_M$) in Gyr.}
    \label{fig:tform}
\end{figure}

To test if we are detecting the equivalent descendant population of the low-mass quiescent galaxies in the aforementioned $z\sim2.5$ overdensity (Sec. \ref{sec:axisratio}), we use spectrophotometric stellar population modeling from \pros{} to examine the SFHs of low-mass quiescent galaxies in Abell 2744 with associated NIRSpec spectroscopy. While we cannot expect to find definitive evidence of a progenitor-descendant link, we can consider all of the stellar population and morphological information together in order to interpret the possible origin of low-mass quiescent galaxies. Figure \ref{fig:tform} shows the mass-weighted formation redshifts of NIRSpec sources (orange stars) and sources from \cite{Cutler2024} associated with the $z\sim2.5$ overdensity (black points) as a function of stellar mass. In general, the Abell 2744 low-mass quiescent galaxies with NIRSpec spectroscopy have much lower formation redshifts ($1<\langle z_\star\rangle_M<3$) relative to those at $z\sim2.5$. Similarly, the bulk photometric-only sample of Abell 2744 low-mass quiescent galaxies also tend to form later. However, there is a significant sub-population of both the NIRSpec and photometry-only sample at $\log(M_\star/M_\odot)\gtrsim8$ that forms at $3<z<5$, which is more in line with the cosmic noon sample.

\begin{figure}
    \centering
    \includegraphics[width=\linewidth]{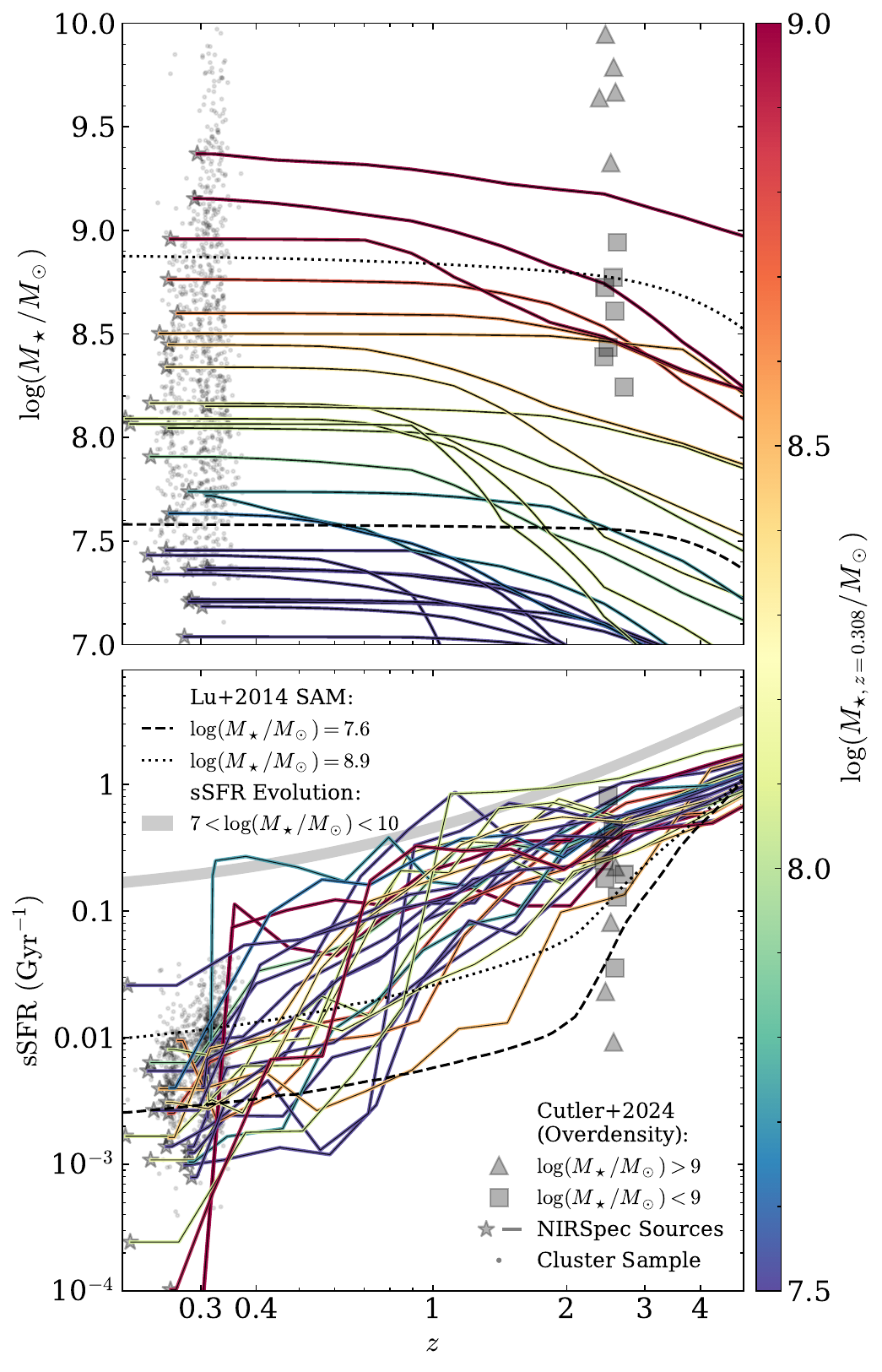}
    \caption{Low-mass quiescent galaxies in Abell 2744 quench star formation and form most of their mass later than those in a $z\sim2.5$ overdensity. In both panels, squares and triangles show the sample of $\log(M_\star/M_\odot)<9$ and $\log(M_\star/M_\odot)>9$ quiescent galaxies from \cite{Cutler2024} that are associated with the $z\sim2.5$ overdensity behind Abell 2744, respectively. Circles show the Abell 2744 sample (with stars indicating NIRSpec sources). Dashed and dotted lines show semi-analytical model mass growth and sSFR as a function of redshift \citep{Lu2014} for $\log(M_\star/M_\odot)=7.6$ and $8.9$ ($M_h=3\times10^{10}h^{-1}$ and $10^{11}h^{-1}M_\odot$), respectively. The shaded grey region shows the redshift evolution of sSFR from \cite{Whitaker2014} between $7<\log(M_\star/M_\odot)<10$, normalized to the \cite{Leja2022} main sequence at $z=0.308$ in order to correct for methodology dependent systematic offsets. In the top panel, solid lines show projected stellar mass growth as a function of redshift for the NIRSpec sources, while the bottom panel shows the spectrophotometrically modeled SFHs of low-mass quiescent galaxies. The solid lines are colored based on their stellar mass at observation ($z=0.308$).}
    \label{fig:trajs}
\end{figure}

Figure \ref{fig:trajs} shows the stellar masses (top) and specific star-formation rates (sSFR, bottom) as a function of redshift for these NIRSpec sources, colored by their stellar masses at the epoch of observation. At $z\sim2.5$, we find that only the $\log(M_\star/M_\odot)\gtrsim8.5$ galaxies have similar stellar masses to the $\log(M_\star/M_\odot)<9$ \cite{Cutler2024} overdensity galaxies (black squares in Fig. \ref{fig:trajs}), in rough agreement with the formation redshift analysis in Figure \ref{fig:tform}. Cosmic noon galaxies at $\log(M_\star/M_\odot)>9$ are clearly too massive to be potential progenitors of the Abell 2744 sample. Both these effects are likely due to completeness: the Abell 2744 sample is less complete at $\log(M_\star/M_\odot)>9$ while the cosmic noon sample of \cite{Cutler2024} is less complete at $\log(M_\star/M_\odot)<9$. In particular, the deeper UNCOVER photometry and crowded cluster environment of the Abell 2744 sample may lead to source detection biases at the more massive end ($\log(M_\star/M_\odot)>9$), as more complex morphologies emerge and sources are contaminated by nearby companions and ICL. We estimate rough completeness limits by comparing the mass distribution of galaxies in our sample to the $0.25<z<0.75$ best-fit quiescent mass function from \cite{Santini2022}, assuming the stellar mass estimates do not suffer from the same incompleteness issues inherent to our size measurements and ignoring cosmic variance. The mass completeness at higher masses is notably suppressed: 56\% at $9.5<\log(M_\star/M_\odot)<10$, compared to 97\% at $8.5<\log(M_\star/M_\odot)<9$. This is likely due to $\log(M\star/M_\odot)>9$ cluster galaxies being missed in our detection step due to segmentation ``shredding''.

Likewise, many of the $\log(M_\star/M_\odot)\gtrsim8$ galaxies also have comparable sSFRs to the cosmic noon sample when traced back to $z\sim2.5$, however they appear to quench at lower redshifts ($z<1.5$). The growth in stellar mass (Fig. \ref{fig:trajs}, top) for most of these galaxies also falls between the \cite{Lu2014} semi-analytic models at $\log(M_\star/M_\odot)=7.6$ and $8.9$ (black dotted and dashed lines). These models predict a transition redshift of $z\sim2$ below which low-mass halos see significantly lower SFRs, a result that \cite{Lu2014} explain via a mix of environmental and physical effects that lead to different SFRs at high redshifts \citep[such as intergalactic medium preheating,][]{Mo2002}. However, we see that this major transition occurs at lower redshifts (Fig. \ref{fig:trajs}, bottom), at least in the Abell 2744 cluster, with sSFRs seeing significant decreases as late as $z\sim0.7$ or even $z\sim0.3$.

\section{Discussion}\label{sec:discuss}
In this work, we observe the structural and star-formation properties of low-mass quiescent galaxies in a low-redshift cluster environment. These galaxies have likely seen significant changes to their structure and star-formation due to environmental effects associated with overdensities. In this section, we explain the potential origins of these structural and stellar population properties within the context of galaxy evolution, low-mass quenching, and effects from cluster environments. 

\subsection{Deviations from Previous Size-Mass Relations}\label{sec:comparison}
The observed differences in size scatter described in Section \ref{sec:size-mass} could be due in part to differing selection methods. For example, \cite{Hamadouche2024} use the $UVJ$ selections of \cite{Williams2009}, which differs from our red-sequence selection (Section \ref{sec:selection}). Moreover, as the results of \cite{Hamadouche2024} are based on PRIMER \citep{Dunlop2021} observations, the (relatively) shallower depth of this survey compared to UNCOVER may bias the sample selection such that they preferentially miss galaxies with lower surface brightnesses, such as potential larger low-mass quiescent galaxies. Moreover, PRIMER is a field survey and thus has a lower density of galaxies at $z\sim0.3$ relative to the targeted Abell 2744 cluster observations. In \cite{Hamadouche2024}, the majority of low-mass quiescent galaxies in the $0.25<z<0.75$ bin are at $z>0.5$. Meanwhile, the \cite{Hamadouche2024} galaxies at $z<0.4$ are notably smaller than the overall population within the $0.25<z<0.75$ bin. This either hints at a potential bias against larger low-mass quiescent galaxies at the lowest redshifts or real systematic differences between field and cluster samples. However, the combined HFF and CANDELS relation from \cite{Nedkova2021}, which likely includes cluster galaxies from HFF, has similar scatter to the Abell 2744 sample at $8<\log(M_\star/M_\odot)<10$, suggesting that our measured scatter is physical.

If the larger size scatter is indeed a physical effect, it could indicate that rich, massive ($\log(M_{\rm vir}/M_\odot)>15$) cluster environments, such as Abell 2744, have a significant population of larger ($R_{\rm e,SMA}>1$ kpc) low-mass quiescent galaxies at $7\lesssim\log(M_\star/M_\odot)\lesssim9$ relative to field environments. This effect could be naturally explained by dynamical interactions within the cluster environment, such as tidal heating, tidal stripping, and ram-pressure stripping, causing low-mass disky galaxies to ``puff up'' or undergo expansion, resulting in a population of ultra-diffuse galaxies \citep[UDGs, e.g.,][]{Safarzadeh2017,Ogiya2018,Carleton2019,Sales2020,Tremmel2020,Benavides2021,Benavides2025,Grishin2021,Jones2021,Junais2022}. While the bulk population in Abell 2744 appears in line with mostly dwarf spheroids (see Section \ref{sec:axisratio}), the higher scatter relative to field studies may be a result of significant contributions from UDGs. UDGs have similar masses and sizes - $\log(M_\star/M_\odot)\lesssim9$ and $R_e>1.5$ kpc, respectively \citep{vanDokkum2015b,Janssens2017,Janssens2019,Lee2017,Carleton2023} - to many of the larger galaxies in our low-mass quiescent galaxy sample. 

In particular, \cite{Janssens2019} find 99 UDGs at $7.7\lesssim\log(M_\star/M_\odot)\lesssim9$ in Abell 2744, suggesting a significant fraction may be present in our sample down to $\log(M_\star/M_\odot)\sim7$. After cross-matching with the \cite{Janssens2019} sample, we find 26 UDGs (37.1\% of the 70 UDGs that overlap with our detection region) are included in the Abell 2744 low-mass quiescent galaxies. The other sources are often detected but assigned a higher redshift in our catalogs, thus it is hard to discern if they are bonafide cluster members or not. These missing UDGs have roughly the same median size ($R_e=2.49$ kpc) as those in our sample, suggesting we are not biased against larger UDGs. Figure \ref{fig:struc} also shows that galaxies in our sample exhibit surface densities as low as the $\Sigma_{\rm e,M_\star}\sim1~{\rm M_\odot~pc^{-2}}$ where Coma UDGs are found \citep{vanDokkum2015b}. The moderate fraction of UDGs found in our sample, combined with the fact that all are located above the measured size-mass relation, indicate that the inclusion of UDGs may explain the higher scatter in Abell 2744. 

\cite{vanderBurg2017} find that UDGs are predominantly found in overdense environments, with their population scaling steeply with halo mass while the population of bright galaxies follows a much shallower relation. Similarly, studies of the Fornax cluster \citep[$\log(M_{\rm vir}/M_\odot)=13.8$,][]{Drinkwater2001} find a population of 9 UDGs \citep{Venhola2017} compared to 288 in Coma \citep{Yagi2016}, the latter cluster of which is more massive by a factor of 20 \citep[$\log(M_{\rm vir}/M_\odot)=15.1$,][]{Ho2022}. Thus, while UDGs are found in both field and cluster environments \citep[e.g.,][]{Zaritsky2023}, including the aforementioned field samples \citep[i.e.,][]{vanderWel2014,Kawinwanichakij2021,Hamadouche2024}, they are more prominent in larger overdensities. Moreover, these studies would not be sensitive to detecting them given the lower surface brightness limits and/or smaller area of the surveys. For example, UDGs typically have surface brightnesses fainter than $\mu_z\sim23~{\rm mag~arcsec^{-2}}$ \citep{Zaritsky2023} whereas, e.g., PRIMER has a $1\sigma$ surface brightness depth of $\mu_{\rm F090W}\sim22.1~{\rm mag~arcsec^{-2}}$ (compared to $23.6~{\rm mag~arcsec^{-2}}$ in UNCOVER). Field studies using PRIMER data \citep[i.e.,][]{Hamadouche2024} are thus unlikely to detect UDGs at $z<0.5$, which suggests that their absence from these samples is due to observational biases and that the measured discrepancies in $\sigma_{\log R_e}$ between the Abell 2744 sample and field samples are likely driven by these biases.

In addition to the lower number densities of Fornax UDGs (relative to more massive clusters), they also are smaller than those from Coma \citep{vanDokkum2015} or Virgo \citep{Mihos2015}, with very few larger than $R_e=2$ kpc (e.g., Fig. 2, \citealt{Munoz2015}; Fig. 8, \citealt{Eigenthaler2018}). Similarly, UDGs associated with group galaxies have less regular morphologies than those in clusters \citep{Merritt2016}, indicating different formation mechanisms than cluster UDGs. The combination of UDGs formed in rich clusters (which may correspond to the previously mentioned ``puffed up'' galaxy population) with smaller, compact dwarf galaxies (those that instead were eroded) - together forming via cluster-specific dynamical processing - sets the cluster size scatter apart from field samples.

\subsection{Intrinsic Shapes of Low-Mass Quiescent Galaxies}
Figure \ref{fig:strucparams} indicates that low-mass quiescent galaxies in Abell 2744 have significantly different axis-ratio distributions from the $1<z<3$ sample from \cite{Cutler2024}, and thus likely have different intrinsic shapes. In order to constrain the intrinsic shapes of the Abell 2744 low-mass quiescent galaxies, we fit their axis-ratio distributions using a three-population model of disky, spheroidal, and elongated (prolate) galaxies. We estimate ellipticity and triaxiality using a least-squares minimization technique that most accurately reproduces our observed axis ratio distributions. From these best-fit parameters, we then generate a sample population as seen from random viewing angles to model the intrinsic axis-ratio distribution. Based on the categories from \cite{vanderWel2014b}, we classify morphology fractions belonging to each intrinsic population that collectively define the axis-ratio distribution of our observed population. 

We find that the Abell 2744 low-mass quiescent galaxies are primarily (76\%) spheroidal, while the low-mass, $t_{50}<500$ Myr quiescent galaxies from \cite{Cutler2024} are composed of only 46\% spheroids. The fraction of spheroids in the Abell 2744 sample is much closer to what we see in the massive, $t_{50}>500$ Myr quiescent galaxies at cosmic noon (78\%). However, this is likely driven by different physical processes: massive galaxy structures are largely impacted by merger and accretion events adding stars, whereas low-mass galaxies are shaped instead by tidal effects from close-encounters that redistribute or even strip away existing stars. The intrinsic shapes of the low-mass galaxies implied by these axis-ratio distributions, along with the low S\'ersic indices, are indicative of a population of predominantly dwarf ellipticals \citep{Graham2003}. The higher axis ratios of Abell 2744 galaxies relative to cosmic noon samples could also be explained by UDG contamination: \cite{Carleton2023} find that the low-stellar-surface-density galaxies, the likely progenitors of $z=0$ UDGs, in the El Gordo cluster ($z=0.87$) have similar S\'ersic indices ($n\lesssim2$) and axis-ratio distributions to the Abell 2744 low-mass quiescent galaxies.

As mentioned previously, only a subset ($\sim16\%$) of the \cite{Cutler2024} cosmic noon sample is likely associated with overdense environments. The overall young ages of the low-mass quenched population from \cite{Cutler2024}, including both field and cluster galaxies, imply that they have experienced only the initial onset of quenching. Multiple quenching mechanisms can act on a galaxy at the same time, making the timescale needed to significantly impact the suppression of star formation the distinguishing factor. As the majority of environmental quenching mechanisms act over longer timescales (excluding ram-pressure stripping), the goal herein is to assess the impact of those effects that are unique to the cluster environment. The full cosmic noon sample is therefore a good morphological comparison point relative to the $z=0.308$ sample, as the latter has had quenching maintained over many Gyr through dynamical interactions with the cluster environment, whereas the former has only just had quenching triggered. Thus, as this comparison can afford to remain agnostic to the root cause of the onset, the focus instead will be on the impact of the specific longer timescale mechanisms that can only occur in dense environments.

\subsection{Low-Mass Quenching}
Environmental mechanisms are often attributed with shutting off star formation in low-mass quiescent galaxies. However, it is currently unclear when these processes become relevant and how this population becomes more prevalent. Recent results suggest that these physical processes could become important as early as cosmic noon: at these redshifts, proto-clusters and other large scale structures form and begin to host a robust ICM capable of efficiently stripping gas \citep[e.g.,][]{Kravtsov2012,DiMascolo2023,Travascio2024,vanMarrwewijk2024,Kiyota2025}. Moreover, several studies find evidence of an upturn in the quiescent galaxy mass function around $\log(M_\star/M_\odot)\sim9$ \citep[e.g.,][]{Santini2022,Weaver2023,Hamadouche2024} as early as $z\sim3$ (though the onset of incompleteness limits precludes more definitive conclusions at $z>2$). In this work, we project the formation histories of low-mass quiescent galaxies in Abell 2744 back to their epoch of formation, and find these to be delayed relative to samples in overdensities found at cosmic noon (Section \ref{sec:form}). Formation redshifts indicate that the majority of galaxies formed at $z<1.5$ (Fig. \ref{fig:tform}), with only the most massive ($\log(M_\star/M_\odot)>8.5$) sources with spectrophotometric modeling having comparable stellar masses to the quiescent population in the $z\sim2.5$ overdensity (Fig. \ref{fig:trajs}).

\begin{figure*}
    \centering
    \includegraphics[width=\linewidth]{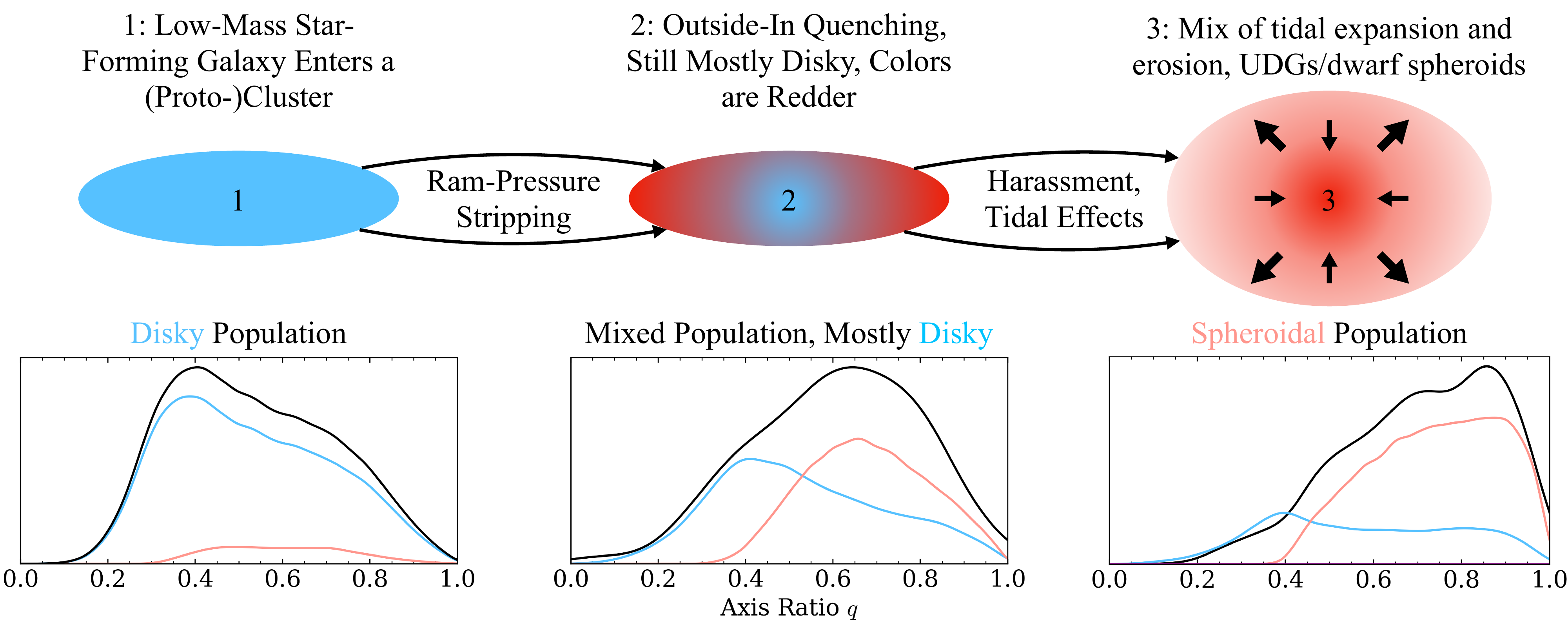}
    \caption{Cartoon showing the evolution from a disky low-mass star-forming galaxy to quiescent dwarf spheroids in a cluster environment. At $z<2$ the hot ICM in (proto-)clusters removes the gas from low-mass star-forming galaxies, leaving a population of predominantly disky, low-mass quiescent galaxies with centrally-concentrated residual star-formation. Further longer-timescale, environmental effects (e.g., dynamical stirring, tidal effects) significantly alter the morphology of low-mass quiescent galaxies in the cluster, either via erosion or expansion. These effects, indicated by black arrows, lead to a mixed population of compact dwarf spheroids and UDGs with a large size scatter but a median size similar to local dwarf galaxies. The bottom panels show the axis-ratio distributions of these three stages broken into intrinsic distributions of disky (blue) and spheroidal (red) galaxies.}
    \label{fig:cartoon}
\end{figure*}

\subsubsection{Low-Mass Quenching at Cosmic Noon}
The formation histories in Section \ref{sec:form} point towards a scenario where the first low-mass quiescent galaxies form \textit{en masse} at $z<1.5$. At the same time, newly discovered populations at cosmic noon ($1<z<3$) are young, with the majority of sources younger than $t_{50}=500$ Myr \citep{Cutler2024}. These young ages suggest that we are seeing the onset of low-mass quenching at $1<z<3$, though in this case it remains unclear what specific mechanisms are responsible. An important disclaimer here is that while the luminosity-weighted ages at cosmic noon \citep{Cutler2024} are unambiguously young by virtue of the blue rest-frame colors of the galaxies, models have not yet reached consensus on the mass-weighted ages inferred for this particular regime, ranging from tens of Myr up to 1 Gyr (\citealt{Kaushal2024}; Gallazzi et al. in prep.). Indeed, these differences between light- and mass-weighted ages may explain existing tensions in reported ages of low-mass quiescent galaxies at cosmic noon; \cite{Pan2025}, using mass-weighted $t_{50}$, find roughly half their low-mass quiescent galaxies at $z\sim2.5$ have $t_{50}>500$ Myr, while the vast majority of the \cite{Cutler2024} galaxies (which are based on light-weighted ages inferred from $UVJ$ colors) are younger than 500 Myr. Regardless, the past SFH of the cosmic noon sample is not particularly relevant here. Rather, the most recent quenching episode, as traced by luminosity-weighted ages, helps guide our interpretation.

Based on the younger ages in \cite{Cutler2024}, and the age of the universe at $z\sim2.5$ ($\sim2.6$ Gyr), environmental quenching mechanisms like starvation, dynamical stirring (i.e. harassment), or tidal effects, which are known to act over longer periods of cosmic time \citep[$>1$ Gyr,][]{Wetzel2013,Hirschmann2014,Wright2019}, are not yet likely to impact this population. This leaves ram-pressure stripping \citep{Gunn1972,Boselli2006} as a potential environmental cause, as it take place on timescales of $0.1-1$ Gyr \cite[e.g.,][]{Samuel2023} and hot ICM has been shown to exist in proto-clusters at these redshifts \citep[e.g.,][]{Kravtsov2012,DiMascolo2023,Travascio2024,vanMarrwewijk2024,Kiyota2025}. However, \cite{Pan2025} analyze the SFHs and sizes of low-mass quiescent galaxies in the $z\sim2.5$ overdensity behind Abell 2744, finding near-identical physical properties and quenching timescales between low-mass galaxies in the field and those in the overdensity. The \cite{Pan2025} results strongly indicate there is no evidence that low-mass quenching is environmentally driven in this  $z\sim2.5$ overdensity, calling into question theories that invoke environment to explain the buildup of low-mass quiescent galaxies at cosmic noon. Moreover, the Abell 2744 low-mass quiescent galaxies, which will undoubtedly have had their quenching maintained by environmental mechanisms, at least in their late-time evolution, tend to form $\sim2$ Gyr later than those in the $z\sim2.5$ overdensity (Fig. \ref{fig:tform}). These results either suggest that a) the initial quenching of low-mass galaxies at $z>2$ is not driven by environment, or b) these galaxies at $z\sim2.5$ are not permanently quenched. 

If ram-pressure stripping is not the cause of low-mass quenching in $z>2$ overdense environments (due to the potential lack of a hot ICM), other mechanisms must be required to quench them at similar epochs to purported high-redshift overdensities. Stellar feedback is often invoked to explain field quenching \citep[e.g.,][]{Fitts2017}. It has also been shown to promote disk formation and correlate with larger sizes and would thus result in an overall flatter size-mass relation \citep[e.g.,][]{Ubler2014,Pillepich2018}. As such, it could explain the non-environmental quenching of low-mass cluster galaxies beyond $z\sim2$. However, stellar feedback may not be able to permanently quench low-mass galaxies as interstellar gas can be replenished by accretion from the circumgalactic medium \citep[e.g.,][]{Tacchella2016}.

The lack of evidence for environmental quenching at $z\sim2.5$, along with the overabundance of young ($t_{50}<500$) low-mass quiescent galaxies at $1<z<3$, strongly hints that $z>2$ low-mass quiescent galaxies are only temporarily quenched. Recent studies have highlighted that high-redshift galaxies caught in a lull within bursty SFHs can be mistaken both photometrically and spectroscopically for ``permanently'' quenched systems (so-called mini-quenched/napping galaxies, e.g., \citealt{Tacchella2016}; \citealt{Angles-Alcazar2017}; \citealt{Faucher-Giguere2018}; \citealt{Ma2018}; \citealt{Looser2023,Looser2024}; \citealt{Sun2023}; \citealt{Dome2024}; \citealt{Gelli2025}; \citealt{Trussler2025}; \citealt{Mintz2025}; G. Khullar et al. in prep.). It is thus possible that some galaxies in the very young population within the high-redshift overdensity \citep{Pan2025} will resume star-formation on moderately short timescales of order 100 Myr or less \citep{Dome2024}.

Indeed, the star-formation and structural properties of these $z\sim2.5$ low-mass quiescent galaxies could in principle be consistent with either scenario, be that recent quenching (from environmental effects or stellar feedback) or temporary quenching via bursty SFHs. These galaxies have sSFRs $0.5-0.7$ dex below the \cite{Whitaker2014} main sequence evolution (grey shading in Fig. \ref{fig:trajs}, bottom), suggesting a population that is either transitioning into quiescence \citep[e.g.,][]{Tacchella2019} or is temporarily quenched. Similarly, the low S\'ersic indices ($n<2$) and flatter axis-ratio distributions (e.g., Fig. \ref{fig:strucparams}) could indicate recent permanent quenching or temporary quiescence. As such, it stands to reason that some fraction of these galaxies are only temporarily quenched, though the exact demographics are challenging to pin down. Future studies could leverage the ages of low-mass quiescent galaxies in overdensities at intermediate redshifts ($1\lesssim z\lesssim1.5$), where (proto-)clusters with hot ICM are more abundant to help distinguish napping quiescent galaxies from environmentally quenched galaxies over a wider baseline.

\begin{figure*}
    \centering
    \includegraphics[width=\linewidth]{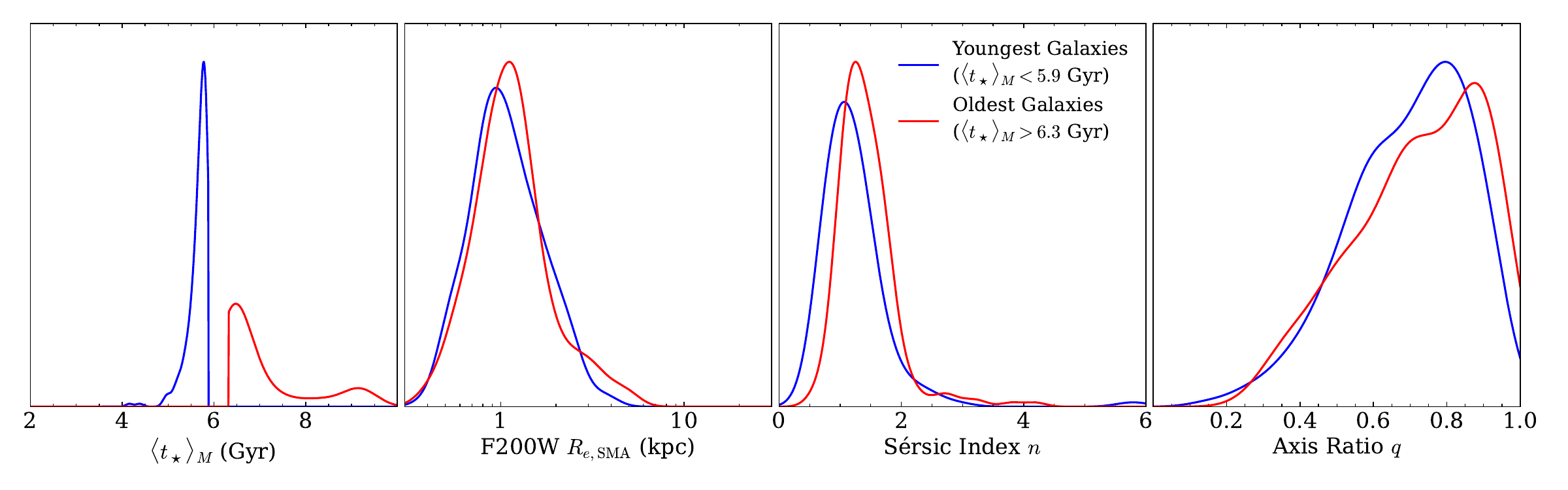}
    \caption{Younger low-mass quiescent cluster galaxies have lower S\'ersic indices and diskier axis-ratio distributions than older sources. Distributions of mass-weighted age ($\langle t_\star\rangle_M$), size ($R_{e,_{\rm SMA}}$), S\'ersic index ($n$), and axis ratio ($q$) for the youngest (blue) and oldest (red) quartiles of low-mass quiescent galaxies in Abell 2744.}
    \label{fig:age_quart}
\end{figure*}

\subsubsection{Connecting Cosmic Noon to Low-Redshift Clusters}
Regardless of whether or not the galaxies associated with the $z\sim2.5$ overdensity are permanently quenched, these galaxies will eventually be impacted by interactions with the cluster environment, leading to permanent quenching and structural changes. These changes must occur in the $\sim10$ Gyr between $z\sim2.5$ and $z\sim0$, since we see dramatic differences in the star-formation and morphological properties of low-mass quiescent galaxies at cosmic noon relative to those observed locally; in the low-redshift ($z=0.308$) cluster, galaxies are much more spheroidal (based on axis-ratio distributions, Fig. \ref{fig:strucparams}) whereas their sizes are roughly similar to high redshift ($1<z<3$) samples for $\log(M_\star/M_\odot)<9$ (Fig. \ref{fig:size-mass}, left). Similarly, the $z\sim2.5$ overdensity sample has sSFRs roughly $1-2$ dex higher than the cluster galaxies in Abell 2744 (Fig. \ref{fig:trajs}, bottom; main sequence shown as benchmark). Moreover, while the low-mass size-mass relation is roughly identical between field and proto-cluster galaxies at $z\sim2.5$ \citep{Pan2025}, at $z=0.308$ we find a significantly larger scatter in Abell 2744 relative to field relations \citep{vanderWel2014,Kawinwanichakij2021,Hamadouche2024}. 

These physical observations allow us to speculate on the evolutionary pathways that connect low-mass galaxies in cosmic noon overdensities to those in low-redshift clusters. The establishment of sufficient amounts of hot ICM in $z\sim2$ proto-clusters allows ram-pressure stripping to remove gas from all but the most interior regions of low-mass galaxies \citep{Gunn1972,Boselli2006}. After this process, the resulting galaxy retains its disky morphology \citep[several Gyr are required to cause significant morphological changes,][]{Marasco2023}, and may have residual, centrally-concentrated star formation \citep{Schaefer2017}. This process is detailed by the transition from stage 1 to 2 in Figure \ref{fig:cartoon}. Assuming the low-mass galaxies in the $z\sim2.5$ overdensity recently underwent permanent quenching, this leftover star formation could explain why they all occupy a ``transitional'' region $0.5-0.7$ dex below the main sequence. The observed disky morphologies and exponential light profiles of low-mass quenched galaxies at cosmic noon is congruent with this picture.

As low-mass quiescent galaxies gradually fall into the cluster, continued interactions with the ICM strip the remaining gas from their centers and further erode the outer stellar regions \citep{Boselli2006}. The $\sim4-7$ Gyr between $z=1-2$ and $z=0.308$ are consistent with the timescales on which recurring dynamical interactions can significantly change the morphology of low-mass galaxies into red, predominantly spheroidal galaxies \citep{Wetzel2013,Hirschmann2014,Wright2019,Marasco2023}. At the same time, dynamical stirring from other cluster galaxies can cause further morphological changes, contributing to the population of dwarf ellipticals \citep{Moore1996}, which have consistent axis-ratio distributions and S\'ersic indices with our low-redshift Abell 2744 sample \citep{Graham2003}. The transition from stage 2 to 3 in Figure \ref{fig:cartoon} highlights this process. Notably, some studies \citep[e.g.,][]{Feldmann2011} argue that morphological changes in cluster environments occur first (as a form of pre-processing) and are later followed by quenching via ram-pressure stripping, a reversal of the process detailed in Figure \ref{fig:cartoon}.

The effects of dynamical stirring and tidal interactions can be erosive or expansive (indicated by the black arrows in stage 3 of Figure \ref{fig:cartoon}), either stripping loosely-bound stellar material and shrinking the size of a galaxy or disturbing ordered motion leading to larger velocity dispersions and sizes. The competition between tidally-driven erosion and expansion may explain the larger size scatter in the size-mass relation relative to field studies. When tidal expansion dominates the result is a population of UDGs found above the size-mass relation, while erosion-dominated galaxies become primarily compact dwarf spheroids. It appears as though the majority of systems become eroded, ending as dwarf spheroids with comparable sizes to local dwarf galaxy samples \citep[e.g.,][]{Eigenthaler2018,Carlsten2021}, though this may be driven by observational biases against very low surface brightness sources. Several studies suggest low-mass galaxies undergo expansion first, while erosive mechanisms occur after tidal heating has already disrupted galaxy structure \citep[e.g.,][]{Gnedin2003,Gnedin2003b,Kazantzidis2011,Paudel2014}. The lowest-mass galaxies may also preferentially lose more stellar material due to their smaller potential wells, reducing their size over time. In this case, the end result of the picture shown in Figure \ref{fig:cartoon} may be mass dependent over sufficiently long timescales, with the lowest-mass sources becoming progressively smaller while slightly more massive sources stay the same on average. 

The smaller population of UDGs in less rich clusters (e.g., Fornax) relative to more massive clusters (e.g., Abell 2744, Coma) may be due to differences in the strength of dynamical processes in these environments. In the most massive clusters, higher speed tidal interactions can inject more energy into stellar orbits, leading to tidally-driven expansion \citep[e.g.,][]{Moore1996,Moore1998,Moore1999,Bialas2015,Safarzadeh2017}. Meanwhile, the lower speeds in less-massive clusters would result in primarily erosion-dominated galaxies, resulting in less UDGs. A similar rationale may explain why UDGs in group environments have irregular morphologies compared to those in clusters \citep{Merritt2016}: the lower velocities in groups allow for more merger interactions to occur, creating elongated and tidally-disrupted UDGs \citep[e.g.,][]{Crnojevic2016,Toloba2016}.

In the most massive cluster environments (such as Abell 2744), the continued accretion of low-mass galaxies could also explain the wide range of axis ratios we find in this work (Fig. \ref{fig:size-mass}) by acting as a kind of progenitor bias: newly-accreted quiescent galaxies at $7<\log(M_\star/M_\odot)<9$ would have larger sizes \citep[e.g.,][]{vanderWel2014} and disky morphologies to those at cosmic noon, while those that have been in the cluster since $z>1$ are more spheroidal but retain similar sizes on average due to the combined effects of tidal erosion and expansion. To test for this ``progenitor bias'', we measure the median mass-weighted age ($\langle t_\star\rangle_M$) above and below the best-fit size-mass relation from Section \ref{sec:structure}. Since local environmentally-quenched low-mass galaxies have older stellar populations than similar-mass field galaxies \citep[e.g.,][]{Trager2000,vanDokkum2003,Thomas2005,Thomas2010,Renzini2006,Paulino-Afonso2020}, which would be the source for newly accreted cluster galaxies, we would expect to see an age gradient across the size-mass relation where larger and predominantly diskier, newly-accreted galaxies have younger ages than smaller, older spheroids. However, while larger galaxies have a median age of $\langle t_\star\rangle_M=6.03$ Gyr that is slightly younger than the $6.14$ Gyr of galaxies below the size-mass relation, the difference is not significant. That said, the difference is a factor of 3 larger if we consider the spectrophotometrically-modeled galaxies only, but there are then small number statistics to contend with. 

In Figure \ref{fig:age_quart}, we compare the mass-weighted age to both S\'ersic index and axis ratio for the youngest (blue) and oldest (red) $\langle t_\star\rangle_M$ quartiles. From this analysis, we find that, despite being roughly identical in size, the youngest sources have smaller S\'ersic indices and a flatter axis-ratio distribution that peaks at a smaller axis ratio ($q\sim0.75$). Though the overall mass-weighted age distribution is narrow, the older galaxies are more spheroidal on average. This is comparable to the results of \cite{Allen2016}, who find that star-forming galaxies at $0.5R_{\rm vir}$ have smaller sizes and larger S\'ersic indices than those at $2R_{\rm vir}$. Moreover, \cite{Allen2016} also find that quiescent galaxies have roughly similar sizes and S\'ersic indices on average regardless of how close they are to the cluster center (and subsequently their age), suggesting that most of the morphological gradient we are seeing is driven by recently quenched/transitioning galaxies. 

\section{Summary}\label{sec:summary}
Understanding how and why low-mass galaxies quench is a crucial but relatively underexplored component of quiescent galaxy formation theory. Massive quiescent galaxies are thought to grow primarily through dissipationless, minor mergers \citep[e.g.,][]{Naab2009,vanDokkum2010,Feldmann2010,vanDokkum2015,Trujillo2011,Cimatti2012,McLure2013,Hamadouche2022}, making these gas-free, low-mass quiescent sources an ideal ``missing link'' in these merger pathways. In this paper, we leverage spectrophotometric data from the JWST UNCOVER and MegaScience programs to study the size, morphology, and formation histories of low-mass quiescent galaxies in Abell 2744. Our main results are listed below:

\begin{enumerate}[(i)]
    \item We measure a similar low-mass ($\log(M_\star/M_\odot)<9$) quiescent size-mass relation compared to local dwarf galaxy studies \citep[e.g.,][]{Eigenthaler2018,Carlsten2021}, as well as a comparable intermediate-mass ($9<\log(M_\star/M_\odot)<10$) relation to field studies \citep[e.g.,][]{Nedkova2021,Cutler2022}. However, we find a larger intrinsic model scatter relative to field studies at similar redshifts \citep[e.g.,][]{vanderWel2014, Kawinwanichakij2021,Hamadouche2024}. If real, a broader size-mass relation in Abell 2744 could be indicative of a mixed population of tidally expanded (ultra-diffuse) and eroded (compact dwarf spheroidal) galaxies. That said, we cannot rule out that these discrepancies are instead driven by incompleteness at $z<0.5$ in the field samples and/or biases at $\log(M_\star/M_\odot)>9$ in our cluster sample.

    \item Low-mass quiescent galaxies in Abell 2744 have axis-ratio distributions consistent with a primarily spheroidal population, which indicates these galaxies are mostly lenticulars and/or dwarf ellipticals \citep{Graham2003}. This represents a stark transition from the disky population of low-mass quiescent galaxies at cosmic noon \citep{Cutler2024}, suggesting that continued effects of the cluster (ram-pressure stripping, dynamical stirring, etc.) cause significant morphological changes between $z\sim2$ and today.

    \item Robust SFHs derived from spectrophotometric SED modeling of Abell 2744 cluster galaxies at $z=0.308$ show that the majority of these galaxies formed much later ($z<1.5$) than low-mass quiescent galaxies associated with a $z\sim2.5$ overdensity. Should an evolutionary link exist, this implies that, at $z>2$, low-mass quenching is not driven primarily by environment.  Moreover, galaxies in the higher-redshift sample may only be temporarily quenched.
\end{enumerate}

Overall, these results allow us to speculate on an evolutionary framework where disky low-mass quiescent galaxies in overdense environments evolve into a population of quenched dwarf spheroids in low-redshift clusters. At $z<2$, the establishment of a hot ICM in early (proto-)cluster environments can permanently quench low-mass galaxies. These galaxies are stripped of their gas by tidal effects and ram-pressure stripping, quenching them from the outside in \citep{Gunn1972,Boselli2006}, but their disky morphologies are largely retained along with some residual, centrally-concentrated star formation \citep{Schaefer2017}. Over the several Gyr separating the initial quenching at $z<2$ and our observations of the cluster at $z=0.308$, low-mass quiescent galaxies undergo morphological changes driven by the continued effects of the cluster \citep{Marasco2023}, creating a population of older spheroids and/or dwarf ellipticals with very low sSFRs.

Further studies utilizing more in-depth SED modeling for the entire sample of low-mass quiescent galaxies in Abell 2744 will be crucial to fully investigate whether we do in fact see younger, newly-quenched disky galaxies entering the cluster sample at later epochs. Spectroscopic redshift surveys, such as ALT \citep{Naidu2024}, and upcoming medium-band surveys, like MINERVA \citep[JWST-GO-7814, PI: Muzzin;][]{Muzzin2025}, will lead to improved redshift constraints and provide an opportunity to find overdensities at higher redshifts, which we can use to directly connect low-mass quenching at cosmic noon to environment. Moreover, future probes into the ages of low-mass quiescent galaxies in $1<z<1.5$ overdensities can help untangle the impact of mini-quenched galaxies on our speculated picture of low-mass quenching.

\section*{Acknowledgments}
This work is based in part on observations made with the NASA/ESA/CSA James Webb Space Telescope obtained from the Space Telescope Science Institute, which is operated by the Association of Universities for Research in Astronomy, Inc., under NASA contract NAS 5–26555. Financial support for programs JWST-GO-1837, JWST-GO-2561, JWST-GO-4111, and JWST-GO-6405 is gratefully acknowledged and is provided by NASA through grants from the Space Telescope Science Institute, which is operated by the Associations of Universities for Research in Astronomy, Incorporated, under NASA contract NAS 5-03127. These observations are associated with programs JWST-GO-2561, JWST-ERS-1324, JWST-DD-2756, JWST-GO-4111, HST-GO-11689, HST-GO-13386, HST-GO/DD-13495, HST-GO-13389, HST-GO-15117, and HST-GO/DD-17231. Some of the data products presented herein were retrieved from the Dawn JWST Archive (DJA). DJA is an initiative of the Cosmic Dawn Center, which is funded by the Danish National Research Foundation under grant No. 140.

\facilities{JWST (NIRCam and NIRSpec), HST (WFC3 and ACS)}

\software{
\textsc{Aperpy} \citep[\url{github.com/astrowhit/aperpy}]{Weaver2023},
\textsc{astrodrizzle} \citep{Gonzaga2012},
\textsc{Astropy} \citep{astropy2013,astropy2018,astropy2022},
\textsc{DYNESTY} \citep{Speagle2020},
\eazy{} \citep{Brammer2008},
\textsc{extinction} \citep{extinction},
\textsc{FSPS} \citep{fsps2009,fsps2010a,fsps2010b},
\galfit{} \citep{Peng2002,Peng2010},
\grizli{} \citep[\url{github.com/gbrammer/grizli}]{grizli},
\textsc{matplotlib} \citep{matplotlib2007},
\textsc{numpy} \citep{numpy2011},
\textsc{Photutils} \citep{photutils2022},
\pros{} \citep{prospector2021},
\textsc{Pypher} \citep{Boucaud2016},
\textsc{Python-FSPS} \citep{pythonfsps2014},
\textsc{Seaborn} \citep{Waskom2021},
\textsc{SciPy} \citep{scipy2020},
\textsc{SEP} \citep{Barbary2016},
\textsc{SFDMap} \citep[\url{github.com/kbarbary/sfdmap}]{Schlegel1998,Schlafly2011},
\textsc{statsmodels} \citep{statsmodels}
}

\bibliographystyle{aasjournal}
\bibliography{references}
\end{document}